\documentstyle[12pt,psfig,epsfig,amsfonts]{article}
\newcommand{\nUV}{\mu^{2}}
\newcommand{\nIR}{\mu^{2}}
\newcommand{\nO}{\mu_{0}^{2}}

\newcounter{saveeqn}
\newcounter{App} %\setcounter{App}{0}
\newcommand{\app}{%
\stepcounter{App}%
\setcounter{saveeqn}{\value{equation}}%
\setcounter{equation}{0}%
\renewcommand{\theequation}{\Alph{App}.\arabic{equation}} }
\newcommand{\appende}{%
\setcounter{equation}{\value{saveeqn}}%
\renewcommand{\theequation}{\arabic{equation}}  }

\begin{document}
%\begin{flushright}
\hspace*{11.4cm }TPR-99-10\\
\hspace*{12.0cm }WUE-ITP-99-014\\
\hspace*{12.0cm }NORDITA-1999/43 HE
%\end{flushright}
\vspace{0.5cm}

\thispagestyle{empty}
\begin{center}
{\Large\bf Pion Form Factor in QCD at Intermediate Momentum Transfers}
\vskip 1.5true cm

{\large\bf
V.~M.~Braun} 
\vskip 0.2true cm
{\it Institut f\"ur Theoretische Physik, Universit\"at
  Regensburg, D-93040 Regensburg, Germany} 

\vskip 1true cm
{\large \bf
A.~Khodjamirian}$^*$
\vskip 0.2true cm
{\it Institut f\"ur Theoretische Physik, Universit\"at
  W\"urzburg, D-97074 W\"urzburg, Germany}

\vskip 1true cm
{\large\bf 
M.~Maul} 
\vskip 0.2true cm
{\it
NORDITA, Blegdamsvej 17, 2100 Copenhagen \O, Denmark}
\end{center}

\vskip 1.0true cm
\begin{abstract}
\noindent

  We present a quantitative analysis of the electromagnetic 
pion form factor in the light-cone sum rule approach, 
including radiative corrections and higher-twist effects. 
 The comparison to the existing  data favors the asymptotic
profile of the pion distribution amplitude and allows to 
estimate  the deviation: 
$(\int \!du/u\,\phi_\pi(u))/(\int \!du/u\,\phi^{\rm as}_\pi(u))
=$ 1.1$\pm$ 0.1 at the scale 1 GeV. 
Special attention is payed to the precise definition and interplay of soft and
hard contributions at intermediate momentum transfer, and to matching
of the sum rule to the perturbative QCD prediction. 
We observe  a strong numerical cancellation between the soft 
(end point) contribution and power suppressed 
hard contributions of higher twist, so that the total nonperturbative
correction to the usual pQCD result turns out to be of order   
  $30\%$ for $Q^2\sim 1$~GeV$^2$.

\end{abstract}

\vspace{3cm}
\noindent $^*${\small \it on leave from Yerevan Physics 
Institute, 375036 Yerevan, Armenia} 

\newpage

\section{Introduction}

There is a clear tendency for QCD-oriented experimental studies
to go for more and more exclusive channels.
All future plans also call for very high luminosity and would 
therefore be perfectly suited for the investigation of exclusive and 
semi-exclusive reactions.  
A problem which hinders all attempts to implement these projects
is the lack of truly quantitative QCD predictions.
It is widely anticipated, see e.g. \cite{Isgur89,Rad91,Kroll,BH94,Zhit95},
that for experimentally accessible 
values of the momentum transfer, the perturbative QCD factorization
for hard exclusive reactions \cite{exclusive} 
receives non-negligible corrections from the so-called soft, 
or end-point contributions, which are essentially nonperturbative.
One practical difficulty is that soft corrections can in many cases 
be mimicked (numerically) by modifying the shape of hadron distribution 
amplitudes. An agreement of perturbative predictions with the data 
cannot, therefore, be used to claim smallness of end-point
effects which have to be estimated independently using a 
certain nonperturbative approach. Creating a systematic framework for a 
study of soft end-point corrections is becoming, thus, increasingly timely.

It has been suggested \cite{BH94} that the soft end-point contribution
to the pion electromagnetic 
form factor can be estimated in a largely model-independent  
way within the framework of light-cone sum rules \cite{LCSR}.
The aim of the present paper is to put this technique on a more quantitative 
footing. To this end we calculate the radiative correction to the 
light-cone sum rule, elaborate on the scale dependence and demonstrate
how the sum rule estimates of the end-point effects can naturally be 
combined with the NLO QCD perturbative calculation. In addition, we 
estimate the twist 6 contribution to the sum rule due to the quark
condensate and find this correction to be small.

The presentation is organized as follows. In Sect.~2 we remind 
basic ideas of the light-cone sum rule approach and derive the 
simplest sum rule. Sect.~3 is devoted to the calculation of the 
radiative correction and separation of soft and hard effects. 
As expected, we find that in the $Q^2\to\infty$ limit the form factor
is dominated by the hard rescattering contribution alone, while 
to the $1/Q^4$ accuracy both soft and hard contributions have to be 
taken into account. Higher-twist corrections to the light-cone sum rule 
are considered in Sect.~4, while Sect.~5 contains the results
of our numerical analysis. Matching of the sum rule with the NLO perturbative
predictions is discussed in Sect.~6. 
Finally, in Sect.~7 we summarize.
The paper contains two appendices where we collect some 
useful but bulky expressions and present the relevant formulae 
for light-cone distributions of the pion.

\section{The method of light-cone sum rules}

The approach is based on the study of the correlation function \cite{CS83}
\begin{equation}
T_{\mu\nu}(p,q) = i\int\! d^4 x \,e^{iqx}
\langle 0| T\{j_\mu^5(0) j_\nu^{\rm em}(x)\} 
| \pi^+(p)\rangle \;,
\label{2:cor}
\end{equation}
where $j_\mu^5 = \bar d\gamma_\mu\gamma_5 u$ and
$j_\nu^{\rm em} = e_u\bar u\gamma_\nu u + e_d\bar d \gamma_\nu d$
is the quark electromagnetic current. With $p^2=m_\pi^2$ and $Q^2=-q^2$ fixed,
the correlation function (\ref{2:cor}) depends on a single invariant 
variable $s=(p-q)^2$. The contribution of the pion intermediate state  
equals
\begin{equation}
T_{\mu\nu}(p,q) =  2if_\pi (p-q)_\mu p_\nu F_\pi(Q^2)\frac{1}
{m_\pi^2 - (p-q)^2}\;, 
\label{2:pion}
\end{equation}
where $f_\pi$ is the pion decay constant and $F_\pi(Q^2)$ is the pion
electromagnetic form factor.
On the other hand, at large negative $(p-q)^2$ and $q^2$ the correlation 
function can be calculated in QCD, in full analogy with the  
$\gamma^*\gamma^*\pi$ transition form factor. A common idea 
of all QCD sum rules is a matching between the 
QCD calculation at Euclidean momenta and the dispersion relation in 
terms of contributions of hadronic states, which allows to estimate the 
hadronic quantity of interest. Specifics of the light-cone sum rules
is how exactly the QCD calculation and matching are done. 
To illustrate this point, 
consider the contribution of the simplest diagram in Fig.~1:  
%
%%%%%%%%%%%%%%%%%%%% FIGURE 1 TREE DIAGRAM %%%%%%%%%%%%%%%%%%%%%%%%%%
\begin{figure}
\centerline{\epsfig{file=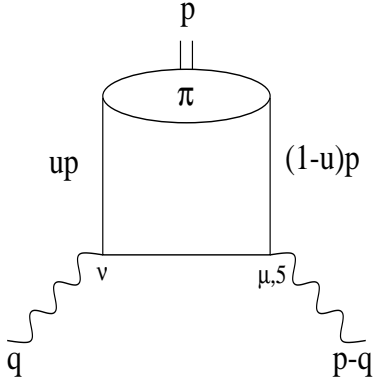, height=5.0cm, width=5.cm}}
\caption[]{\small
 The tree-level contribution to the correlation function
 in Eq.~(\protect{\ref{2:cor}}).
   }
\end{figure}
%%%%%%%%%%%%%%%%%%%% END OF FIGURE 1 %%%%%%%%%%%%%%%%%%%%%%%%%%
%
\begin{equation}
T_{\mu\nu} =  \frac{1}{2\pi^2}\int\!d^4x\,\frac{e^{iqx}}{x^4}
\langle 0|[
e_u \bar d(0)\gamma_\mu\not\!x\gamma_\nu\gamma_5 u(x)
- e_d\bar d(x) \gamma_\nu\not\!x\gamma_\mu\gamma_5 u(0)] 
| \pi^+(p)\rangle \,.
\label{2:ex1}
\end{equation}
Expansion of the remaining nonlocal matrix elements around the 
middle point in a formal Taylor series 
generates the Wilson operator-product expansion 
in contributions of local operators of increasing dimension
\begin{equation}
   O^n_{\mu,\mu_1,\mu_2,\ldots,\mu_n} = \bar d(0)\gamma_{\mu}\gamma_5 
    \,i\!\!\stackrel{\leftrightarrow}{D}_{\mu_1}\ldots
    \,i\!\!\stackrel{\leftrightarrow}{D}_{\mu_n} u(0)\,,
\label{2:ex2}
\end{equation}
$D_\mu=\partial_\mu -igA_\mu$ being  the covariant derivative
and $\stackrel{\leftrightarrow}{D}_\mu \,= 
\,\stackrel{\leftarrow}{D}_\mu\!-
\!\stackrel{\rightarrow}{D}_\mu$.
Restricting ourselves for the moment to operators of the lowest 
twist (highest Lorentz spin) we consider the relevant reduced 
matrix elements 
\begin{equation}
  x_{\mu_1}\ldots x_{\mu_n}\langle 0| O^n_{\mu,\mu_1,\mu_2,\ldots,\mu_n} 
| \pi^+(p)\rangle =
     if_\pi p_\mu (px)^n \langle\!\langle O_n \rangle\!\rangle +\ldots \,.  
\label{2:ex3}
\end{equation}
They are related, as first found in \cite{exclusive}, to the moments of the 
pion distribution amplitude
\begin{eqnarray}
\langle 0| \bar d(0) \gamma_{\mu} \gamma_5 u(x) | \pi^+(p) \rangle 
&=& i p_{\mu} f_\pi \int\limits_0^1 du e^{-iupx}
\varphi_{\pi}(u,\mu^2\sim x^{-2})\,,
\nonumber \\
  \langle\!\langle O_n \rangle\!\rangle &=& 
              \int\limits_0^1\!du\, (1-2u)^n\varphi_{\pi}(u)\,. 
\label{2:ex4}
\end{eqnarray}
Substituting Eq.~(\ref{2:ex3}) in the expansion of (\ref{2:ex1}) and 
integrating over $x$, we obtain for the contribution of 
Fig.~1\footnote{The terms with odd $n$ vanish because of G-parity.}
\begin{eqnarray}
   T_{\mu\nu} & = & \frac{2if_\pi p_\mu p_\nu}{Q^2-s} 
\Bigg\{1+\frac{2Q^2}{Q^2-s}\sum_{n=2,4,\ldots}
\langle\!\langle O_n \rangle\!\rangle
\left(\frac{2Q^2}{Q^2-s}-1\right)^{n-1}\Bigg\}
\nonumber\\
  && {}+ \mbox{\rm other Lorentz structures},
\label{2:ex5}
\end{eqnarray}
where $s=(p-q)^2$. To construct a sum rule, we make the Borel 
transformation 
\begin{eqnarray}
 \frac{1}{m_\pi^2-s} &\longrightarrow& \exp[-m_\pi^2/M^2]\,,
\nonumber\\
 \frac{1}{(Q^2-s)^n} &\longrightarrow& \frac{1}{(M^2)^{n-1}
\Gamma(n)}\exp[-Q^2/M^2]\,,
\label{2:Borel}
\end{eqnarray}
introducing a new variable $M^2$ (the Borel parameter), 
and equating the Borel-transformed versions of Eqs.~(\ref{2:pion})
and (\ref{2:ex5}). For simplicity we neglect
the continuum subtraction here. Neglecting the  pion mass,
 the result reads
\begin{equation}
  F_\pi(Q^2) = e^{-Q^2/M^2}\Bigg\{1+\sum_{n=2,4,\ldots}
\langle\!\langle O_n \rangle\!\rangle
\sum_{k=1}^n \left(\begin{array}{c} n\!-\!1\\k\!-\!1\end{array}
\right)\frac{1}{\Gamma(k+1)}
\left(\frac{-2Q^2}{M^2}\right)^k\Bigg\}.
\label{2:bad} 
\end{equation}    
This sum rule is, however, completely unsatisfactory!

Indeed, QCD sum rules are generally expected to hold in a certain 
interval of values of the Borel parameter, such that contributions
of both higher resonances and higher orders of the OPE are 
simultaneously suppressed. 
It is easy to see that in the present situation these two conditions
are contradictory, unless $Q^2$ is sufficiently small.
Indeed, on the one hand, one has to keep $M^2$ small, of order $1-2$~GeV$^2$,
to suppress the contribution of, e.g., the $a_1$-meson intermediate state.
On the other hand, for a fixed $M^2$ the higher order terms on the r.h.s. of 
the sum rule are enhanced by factors $(Q^2)^k$ and for $Q^2>M^2$ the  
OPE expansion breaks down.

An escape suggested in  \cite{LCSR} is to avoid the Wilson short-distance
expansion altogether and write the answer for the diagram in Fig.~1 directly
in terms of the pion distribution amplitude. The expansion parameter 
then becomes the {\em twist} of the operators rather than 
their dimension. 
Using Eq.~(\ref{2:ex1}) and the definition of the pion distribution 
amplitude in Eq.~(\ref{2:ex4}) we obtain to leading twist 
accuracy, instead of Eq.~(\ref{2:ex5}), a compact expression
\begin{equation}
  T_{\mu\nu} =  2if_\pi p_\mu p_\nu \int\limits_0^1 \! du\! 
   \frac{u\varphi_\pi(u)}{\bar u Q^2 -us}+\ldots\,,
\label{2:ex6}
\end{equation}   
where $\bar u =1-u~$. Making, once again, the Borel transformation, we get the 
simplest {\em light-cone sum rule} \cite{BH94}
\begin{equation}
F_{\pi}(Q^2) = \int\limits_0^1 \!du\, \varphi_{\pi}(u) 
\exp \left( - \frac{\bar uQ^2}{u M^2} \right). 
\label{2:SR1}
\end{equation} 
This sum rule is 
perfectly well behaved at $Q^2\to\infty$ and
it is instructive to trace how the above-mentioned difficulties
of the standard approach have been resolved.  
Because of the strong exponential suppression factor, the important 
region of integration over the momentum fraction variable $u$ gets 
shifted, in the large-$Q^2$ limit,  
to the end-point region $1-u \sim M^2/Q^2$. In this 
regime, the virtuality of the quark (the denominator in Eq.~(\ref{2:ex6}))
remains all the time of order $M^2$, as $Q^2\to\infty$. 
The deficiency of the short-distance expansion 
is now clearly seen
as originating from the wrong expansion parameter $(Q^2-s)/2$
(cf. Eq.~(\ref{2:ex5})) 
corresponding, effectively,  to the expansion around 
the symmetric point $u=1/2$
\footnote{A similar deficiency of the short-distance expansion in the 
case of heavy-to-light correlation functions 
is demonstrated in  \cite{BBKR}.}.

To be somewhat more quantitative, we have to make the usual 
continuum subtraction. This is trivial in the case at hand, since 
the expression  (\ref{2:ex6}) is easily converted to the form of a 
dispersion integral over $s=(p-q)^2$. All we have to do is to truncate 
this integral at a certain threshold $s_0$, called the interval
of duality. The result \cite{BH94} is that the integration over the 
momentum fraction is cut from below at the value 
\begin{equation}
   u_0 = Q^2/(s_0+Q^2)\,.
\label{2:u0}
\end{equation}
In addition, the pion distribution amplitude has to be taken 
at the scale corresponding to the quark virtuality
\begin{equation}
    \mu_u^2 = \bar u Q^2 + u M^2\,.   
\label{2:scale}
\end{equation}  
Implementing these small improvements, 
we obtain the leading-twist leading-order light-cone
sum rule \cite{BH94}
\begin{equation}
F_{\pi}(Q^2) = \int\limits_{u_0}^1 \!du\, \varphi_{\pi}(u,\mu_u) 
\exp \left( - \frac{\bar uQ^2}{u M^2} \right). 
\label{2:SR2}
\end{equation} 
The crucial advantage of the light-cone sum rule approach is that 
it allows to incorporate the information on the end-point behavior 
of the pion distribution amplitude $\varphi_\pi(u) \stackrel{u\to1}{\sim}
1-u$. In the limit $Q^2\to\infty$ the integration
region in Eq.~(\ref{2:SR2}) shrinks to a point $u=1$ so that one obtains
\begin{equation}
  F_\pi(Q^2) \sim \frac{\varphi_\pi'(0,\mu^2\sim M^2)}{Q^4}
   \int\limits_0^{s_0}\!sds\,e^{-s/M^2}\,,
\label{2:SR2limit}
\end{equation} 
where $\varphi_\pi'(0) \equiv (d/du)\varphi_\pi(u)|_{u\to0}
= -\varphi_\pi'(1)$. 
The Borel variable $M^2$ corresponds to the (inverse)
distance at which the matching is done between the parton 
and hadron representations.

The expressions in Eqs.~(\ref{2:SR2}), (\ref{2:SR2limit}) present a 
typical `soft' or `end-point' contribution to the pion form factor 
which is sensitive to the pion wave function at a {\em low} 
normalization point and comes from large transverse distances 
of order $b\sim s_0^{-1/2}$.

To illustrate this point, write the four-dimensional integration in
Eq.~(\ref{2:cor}) as a product of two two-dimensional integrations
in longitudinal and transverse (to $p$ and $q$) coordinates. 
Leaving the transverse integration intact, a short calculation    
gives for the r.h.s. of Eq.~(\ref{2:SR1})
\begin{equation}
  \int \!d^2b\! \int\limits_0^1 \!du\, \varphi_{\pi}(u) \frac{uM^2}{4\pi} 
\exp \left( -\frac14 uM^2b^2-\frac{\bar uQ^2}{u M^2} \right). 
\label{2:b}  
\end{equation}
%
%
%%%%%%%%%%%%%%%%%%%% FIGURE 2 TRANSVERSE SEPARATION  %%%%%%%%%%%%%%%%%%%%%%
\begin{figure}[tb]
\vspace{1cm}
%\centerline{\epsfig{file=bdistribution.eps, height=5.5cm, width=7.cm}}
\centerline{\epsfig{file=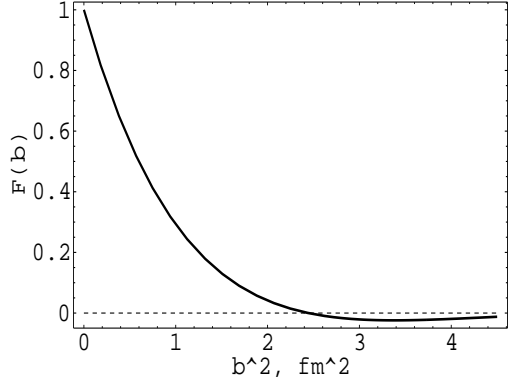, height=5.5cm, width=7.cm}}
\caption{\small
 The transverse-distance separation between the quark and 
 the antiquark in the leading-order light-cone 
 sum rule (\protect{\ref{2:SR2}}) in the large $Q^2$ limit for 
 typical values of the sum rule parameters $s_0=0.7$~GeV$^2$ and
 $M^2= 1.0$ GeV$^2$.}
\end{figure}
%%%%%%%%%%%%%%%%%%%% END OF FIGURE b-separation %%%%%%%%%%%%%%%%%%%%%%%%%%
%
%
%
The distribution of transverse distances in the diagram in Fig.~1
is, thus, gaussian, with the average transverse size 
$\langle b^2\rangle = 4/(uM^2)$
controlled by the value of the Borel parameter. One also sees that
the scale of the distribution amplitude in Eqs.~(\ref{2:scale}), 
(\ref{2:SR2}) is determined  by the weighted average 
of  the momentum transfer $Q^2$ and the (inverse) transverse distance 
between the quarks, as expected on general grounds \cite{LS92}.  

Including the continuum subtraction modifies this 
distribution rather significantly as  the small-$b$ region is 
dominated by high-mass excitations and gets suppressed. 
After some algebra we obtain the sum rule
equivalent to Eq.~(\ref{2:SR2}) but 
with an explicit separation of different transverse distances:
\begin{eqnarray}
F_\pi(Q^2) &=& \frac{1}{4\pi}\int d^2b \int\limits_{u_0}^1\!du\,
\varphi_\pi(u)\,
e^{ -\bar uQ^2/(u M^2)}\!\!
\int\limits_{0}^{u s_0-\bar u Q^2}\! \!\!dt\, e^{-t/(uM^2)} 
\, J_0(\sqrt{b^2 t})
\nonumber\\
&\stackrel{Q^2\to\infty}{\longrightarrow}&
\frac{\varphi'_\pi(0)}{4\pi Q^4}\int d^2b
\int\limits_0^{s_0}
\!ds\,e^{-s/M^2}\!\int\limits_0^sdt (s-t)J_0(\sqrt{b^2 t})\,,
\label{2:Bessel}
\end{eqnarray}
where $J_0 $ is the Bessel function. 
The resulting transverse-distance distribution (normalized to unity at $b=0$)
is shown in Fig.~2. The dependence on both $Q^2$ and 
the Borel parameter is actually very 
weak and the overall scale of transverse distances is determined 
almost entirely by the value of the continuum threshold. 
Because of this, for $M^2\gg s_0$ the pion 
distribution amplitude has to be taken at the scale 
$\mu^2\sim s_0$, rather than at 
$\mu^2 \sim M^2 $ ~\footnote{It can be shown that this change of scale 
takes into
account the continuum subtraction in the running coupling, cf. \cite{BKY85}.}.
The width of the $b^2$-distribution in Fig.~2 should be compared 
with the electromagnetic pion diameter squared\  $(2R_\pi^{\rm em})^2
\sim 2$~fm$^2$.

\section{Radiative corrections}
\subsection{General case}

In order to improve the accuracy of the light-cone sum rule 
(\ref{2:SR2}), one has to calculate the $O(\alpha_s)$ radiative corrections 
to the leading-order correlation function (\ref{2:ex6}). The corresponding 
Feynman diagrams are shown in Fig.~3.
The calculation is straightforward, albeit tedious, 
and technically similar to the calculation
of the radiative correction to the $\gamma^*\gamma^*\pi $ transition 
form factor for different photon virtualities \cite{pigg}.
We handle ultraviolet and infrared collinear divergences by
dimensional regularization in the $\overline{\mbox{MS}}$ scheme.
Due to the fact that the diagrams contain two $\gamma_5$ matrices
(one from the axial-vector vertex and one from the pion projection) 
there is no $\gamma_5$ ambiguity. We have also checked that 
the collinear divergences are absorbed in the definition of the 
scale-dependent pion distribution amplitude $\varphi_\pi$.  
%
%
%%%%%%%%%%%%%%%%%%%% FIGURE 3 radiative corr. %%%%%%%%%%%%%%%%%%%%%%%%%%
\begin{figure}
\centerline{\epsfig{file=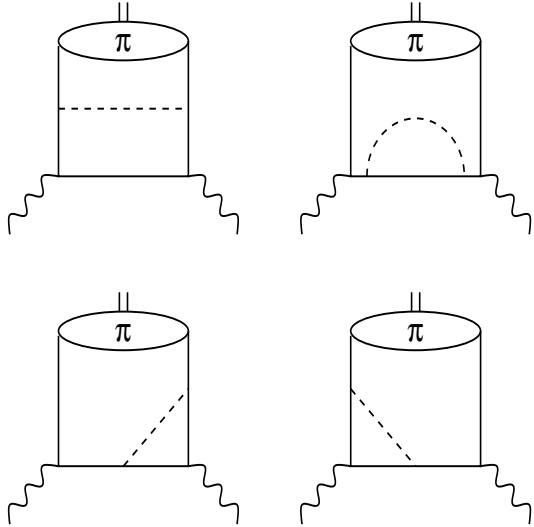, height=7.0cm, width=7.cm}}
\caption[]{\small
 The radiative corrections to the correlation function  
 in Eq.~(\protect{\ref{2:cor}}). Dashed lines denote virtual gluons.
   }
\end{figure}
%%%%%%%%%%%%%%%%%%%% END OF FIGURE 3 %%%%%%%%%%%%%%%%%%%%%%%%%%
%
%
Our result for the twist 2 part of the correlation function 
(\ref{2:cor}) to $O(\alpha_s)$ 
can be represented in a form of convolution of $\varphi_\pi$ 
with the hard scattering amplitude 
\begin{equation}
T^{(2)}_{\mu\nu} =  2if_\pi p_\mu p_\nu \int\limits_0^1 \! du ~\varphi_\pi(u,\mu)
\left\{ H_0(Q^2,s,u)+ \frac{\alpha_sC_F}{4\pi}H_1(Q^2,s,u,\mu) \right\},  
\end{equation}
where the leading-order result was already given in Eq.~(\ref{2:ex6}): 
\begin{equation}
H_0(Q^2,s,u)= \frac{u}{\bar u Q^2 -us}\,,
\end{equation}
and the radiative correction to the hard scattering amplitude equals
\begin{eqnarray}
H_1(Q^2\!\!,s,u,\mu)&=& 
\frac{Q^2}{u \bar u(Q^2+s)^3(\bar u Q^2 -us)}\Bigg[ -9u \bar{u}(Q^2+s)^2   
\mbox{\hspace{5.7cm}}\hfill\break
\nonumber\\
&-&{}\!
\left[Q^4\bar{u}(3\bar{u}-2)+Q^2s(5-6u\bar{u})+s^2u(3u-2)\right]
\ln\left(\frac{\bar u Q^2 -us}{\mu^2}\right)
\nonumber\\
&+&{} \!            
\left[ Q^4\bar{u}u+Q^2s(1+2u\bar{u})+s^2\bar{u}u\right]
\ln^2\left(\frac{\bar u Q^2 -us}{\mu^2}\right)
\nonumber\\
&+&{} \!
u\left[-2Q^4\bar{u}+5Q^2s+s^2(1+2\bar{u})\right] \ln \frac{s}{\mu^2} 
-us\left[ Q^2(1+\bar{u})+s\bar{u} \right]\ln^2\frac{s}{\mu^2}
\nonumber\\  
&+&{} \!
\bar{u}\left[Q^4(1+2u)+5Q^2s-2s^2u\right]\ln\frac{Q^2}{\mu^2}
-\bar{u}Q^2\left[Q^2u +s(1+u)\right]\ln^2\frac{Q^2}{\mu^2}\Bigg].
\label{3:H1}
\end{eqnarray}
To obtain the radiative correction to the 
light-cone sum rule for $F_{\pi}$, one has  to 
calculate the imaginary part of $H_1$ in the variable $s$.   
The resulting expression is presented in Appendix A. 
After continuum subtraction and Borel transformation,
the sum rule reads: 
\begin{equation}
 F^{(2)}_{\pi}(Q^2) = \int\limits_{0}^1 \!du\, \varphi_{\pi}(u, \mu) 
\Bigg[ \Theta(u-u_0){ \cal F}^{(2)}_{\rm soft}(u,M^2,s_0)+
       \Theta(u_0-u){ \cal F}^{(2)}_{\rm hard}(u,M^2,s_0)\Bigg],
\label{3:NLO}
\end{equation}
where
\begin{eqnarray}
\lefteqn{
{\cal F}^{(2)}_{\rm soft}(u,M^2,s_0)=
}
\nonumber\\ 
&=&\exp \left( - \frac{\bar uQ^2}{u M^2} \right)
\left \{ 1 + \frac{\alpha_s}{4\pi}C_F \Bigg[
-9 + \frac{\pi^2}{3} + 3\ln \frac{Q^2}{\mu^2} 
+ 3\ln\frac{\bar u Q^2}{u \mu^2}
-\ln^2 \frac{Q^2}{\mu^2} - \ln^2 \frac{\bar u Q^2}{u \mu^2}\Bigg]\right\}
\nonumber \\
&& +\frac{\alpha_s}{4\pi}C_F \Bigg\{
\int\limits_{\bar u Q^2/u}^{s_0} \frac{ds\, Q^2 e^{-s/M^2}}{u (Q^2+s)^3}
\Bigg[ 5s +Q^2\left(1+2\ln \frac{-\rho}{\mu^2}\right)
+2\left(\frac{Q^2}{\bar u}+ s \right)\ln \frac{-\rho}{s}
\nonumber \\
&& + \frac{2Q^2}{u}\left(\frac{Q^2+s}{s} +\frac{2M^2+Q^2+s}{M^2}
\ln\frac{-\rho}{s} \right)\ln\frac{-\rho}{\mu^2} \Bigg]
\nonumber
 \\
&& +  \int\limits_0^{\bar u Q^2/u} \frac{ds\, Q^2 e^{-s/M^2}}{u \bar u (Q^2+s)^3}
\Bigg[ 2u \left(Q^2-s+s\ln\frac{s}{\mu^2}\right)
+\Bigg(-Q^2 + 5s+2(Q^2-s)\ln\frac{s}{\mu^2} 
\nonumber
\\
&&
-\frac{s(Q^2+s)}{M^2} \left(-3+2 \ln\frac{s}{\mu^2} \right)\Bigg)
\ln\frac{\rho}{\mu^2}\Bigg] 
+ 2\frac{u_0^2}{u^2} e^{-s_0/M^2}
\ln\frac{-\rho_0}{\mu^2}\ln \frac{u-u_0}{\bar u_0} 
\Bigg \}\,,
\label{3:soft2}
\end{eqnarray}
and
\begin{eqnarray}
\lefteqn{{\cal F}^{(2)}_{\rm hard}(u,M^2,s_0) =}
\nonumber\\
&=& \frac{\alpha_s}{4\pi}C_F \Bigg\{
\int\limits_0^{s_0}\frac{ds\, Q^2e^{-s/M^2}}{\bar u(Q^2+s)^3} 
\Bigg[ 2\left(Q^2 -s +s \ln\frac{s}{\mu^2}\right)
+\frac{1}{u }\Bigg(
-Q^2 + 5s+2(Q^2-s)\ln\frac{s}{\mu^2}
\nonumber
\\
&&-\frac{s (Q^2+s)}{M^2}
\left(-3+2 \ln\frac{s}{\mu^2}\right) \Bigg)\ln\frac{\rho}{\mu^2}\Bigg] 
- \frac{u_0\bar u_0}{u \bar u}e^{-s_0/M^2}
\left(2\ln\frac{s_0}{\mu^2}-3\right)
\ln\frac{\rho_0}{\mu^2} \Bigg\}.   
\label{3:hard2}
\end{eqnarray}
Here $\rho = \bar u Q^2 - us$ and $\rho_0 = \bar u Q^2 - us_0= (1-u/u_0)Q^2$. 
The superscript `$^{(2)}$' indicates the leading twist 2 contribution.
Higher twist terms will be added in the next section. 
We interpret the parts of Eq.~(\ref{3:NLO})
with ${\cal F}^{(2)}_{\rm hard}(u,M^2,s_0)$ and
${\cal F}^{(2)}_{\rm soft}(u,M^2,s_0)$ as ``hard'' and ``soft'' 
contributions to the pion form factor, respectively, defined with the
explicit cutoff in the momentum fraction $u=u_0\sim 1-s_0/Q^2$. 
This separation  will be discussed in detail below. 

\subsection{Study case: asymptotic distribution amplitude}

For the asymptotic shape of the pion distribution amplitude
$\varphi_\pi^{\rm as}(u) = 6u(1-u)$ the momentum-fraction integration
in the radiative correction can easily be done analytically, 
with the simple result
\begin{equation}
F_{\pi}^{\rm as}(Q^2) = 6 \int\limits_0^{s_0}\!ds\,e^{-s/M^2}
\frac{s\,Q^4}{(s+Q^2)^4}\left\{
 1+\frac{\alpha_s C_F}{4\pi}\left[\frac{\pi^2}{3}-6-\ln^2\frac{Q^2}{s} +
     \frac{s}{Q^2}+\frac{Q^2}{s}\right]\right\}.   
\label{3:SRas}
\end{equation} 
All the scale-dependent logarithmic terms cancel in this case, as expected.
For large $Q^2\gg s_0 $ one can expand the sum rule (\ref{3:SRas}) 
in powers of $1/Q^2$: 
\begin{eqnarray}
F_{\pi}^{\rm as}(Q^2) &=& 
\frac{3\alpha_s C_F}{2\pi Q^2} \int\limits_0^{s_0}\!ds\,e^{-s/M^2}
\nonumber\\
&&{}+ \frac{6}{Q^4}\int\limits_0^{s_0}\!ds\,s\,e^{-s/M^2}
\left\{
 1-\frac{\alpha_s C_F}{4\pi}\left[10-\frac{\pi^2}{3} + \ln^2\frac{Q^2}{s}
    \right]\right\} + O(1/Q^6)\,.   
\label{3:expand}
\end{eqnarray}
To interpret the leading term, we notice that the integral 
$\int\limits_0^{s_0}\!ds\,e^{-s/M^2}$ can be related to the
 pion decay constant through the QCD sum rule \cite{SVZ}
\begin{equation}
 f_\pi^2 = \frac{1}{4\pi^2}\int\limits_0^{s_0}\!ds\,e^{-s/M^2}
\left(1+\frac{\alpha_s}{\pi}\right) +
  \frac{\langle0|\alpha_s/\pi G^2|0\rangle}{12 M^2}
+\frac{176}{81M^4}\pi\alpha_s\langle \bar q q\rangle^2+\ldots \,.
\label{3:SRfpi}
\end{equation} 
The perturbative correction and the  
gluon- and quark-condensate contributions  
involve an extra power of $\alpha_s$ and are absent, therefore,
in our approximation. Substituting 
$\int\limits_0^{s_0}\!ds\,e^{-s/M^2}\to 4\pi^2f_\pi^2$, we obtain
\begin{equation}
  F_{\pi}^{\rm as}(Q^2) \longrightarrow \frac{8\pi\alpha_s f_\pi^2}{Q^2}\,,
\end{equation}
which coincides with the classical result \cite{exclusive}.

It is easy to see that the $O(1/Q^2)$ contribution to the form factor
comes entirely from the term which we have identified as ``hard'',
while all power suppressed  corrections involve both hard 
and soft contributions.
In particular, we obtain to $O(1/Q^4)$ accuracy
\begin{eqnarray}
\lefteqn{ F_{\pi}^{\rm as, hard}(Q^2) \equiv 
\int\limits_{0}^{u_0} \!du\, \varphi_{\pi}^{\rm as}(u) 
      \,{ \cal F}^{(2)}_{\rm hard}(u,M^2,s_0) =}
\nonumber\\
&=&\frac{3\alpha_s C_F}{2\pi Q^2} \int\limits_0^{s_0}\!ds\,e^{-s/M^2}
\left\{1-\frac{s}{Q^2}\left[
 1 +2\frac{s_0}{s} + \ln \frac{s}{\mu^2} +
\Big(3-2\ln \frac{s}{\mu^2}\Big)
\ln \frac{Q^2}{s_0-s}\right]\right\}. 
\label{3:hard_expand}
\end{eqnarray} 
The soft part is then 
just what is left when this hard contribution is subtracted from the total 
result in Eq.~(\ref{3:expand}). Notice that the separation of soft and hard 
contributions  depends on the collinear factorization scale, even 
for the asymptotic distribution amplitude. We will elaborate on this 
dependence in what follows.

In the local duality limit $M^2\to\infty$ one obtains 
\begin{equation}
 F_{\pi}^{\rm as, hard}(Q^2) = 
\frac{3\alpha_s C_F}{2\pi Q^2}s_0\left\{1-\frac{s_0}{Q^2}
\left[\frac{13}{2}-\frac{\pi^2}{6}+ \ln\frac{Q^2}{s_0}\ln\frac{\mu^2}{s_0}
+\ln\frac{\mu^2}{s_0}+2\ln\frac{Q^2}{s_0}\right]\right\} \,,
\label{3:localhard}
\end{equation} 
and
\begin{equation}
 F_{\pi}^{\rm as, soft}(Q^2) = \frac{3s_0^2}{Q^4}+
 \frac{3\alpha_s C_F}{4\pi Q^4}s_0^2
\left\{\frac52 +\ln^2\frac{\mu^2}{s_0}-\ln^2\frac{Q^2}{\mu^2}+
2\ln\frac{\mu^2}{s_0}+3 \ln\frac{Q^2}{s_0}\right\},
\label{3:localsoft}
\end{equation}
where $s_0\simeq 4\pi^2 f_\pi^2$, cf. Eq.~(\ref{3:SRfpi}).
Note that the $O(1/Q^4)$ hard contribution is large and negative, while 
the soft radiative correction $O(\alpha_s/Q^4)$ is positive, unless 
$Q^2 \gg s_0,\mu^2$. This implies  considerable 
cancellations in the sum of the soft and hard contributions so that 
in order to make their  separation  physically meaningful
 one must assume a low value of the factorization scale 
$\mu^2\sim s_0$.\footnote{It is easy to see that for $\mu^2=Q^2$ 
there are double-logarithmic contributions $\sim \ln^2(Q^2/s_0)$ to   
Eqs.~(\ref{3:localhard}) and (\ref{3:localsoft}) which have opposite sign and 
partially cancel in the sum.}

Finally, notice the double-logarithmic contribution $\sim \ln^2 Q^2/s$ 
in Eq.~(\ref{3:SRas}) which is reminiscent of the Sudakov logarithms 
discussed in \cite{LS92}. A typical size
of these corrections is of order $\ln^2 Q^2/s_0$ which for $s_0\sim
0.7-0.8$~GeV$^2$ and $Q^2\sim 1-10$~GeV$^2$
is much less than $\ln^2 Q^2/\Lambda_{\rm QCD}^2$
with $\Lambda_{\rm QCD}\sim 200$~MeV, as usually assumed. 
For this reason, exponentiation of Sudakov  corrections is numerically 
not important in the present approach.    

\subsection{The $1/Q^2$ expansion in general case}

For a generic pion distribution amplitude the NLO light-cone sum rule
 (\ref{3:NLO}) again simplifies considerably upon the 
expansion in powers of $1/Q^2$. We obtain to $O(1/Q^4)$ accuracy
\begin{eqnarray}
F_{\pi} (Q^2) & = &
\frac{\alpha_s}{2\pi}C_F\int\limits_0^{s_0} \frac{ds\,  e^{-s/{M^2}}}{Q^2} 
\int\limits_0^1 du\, \frac{\varphi_{\pi}(u)}{\bar u}
\nonumber \\&&{}+
\varphi_{\pi}'(0) \int\limits_0^{s_0} \frac{ds\, s\, e^{-s/{M^2}}}{Q^4}
\Bigg[ 1 + \frac{\alpha_s}{4\pi}C_F \Bigg(-9 + \frac{1}{3}\pi^2 + \ln
\frac{s}{\mu^2} - \ln^2\frac{s}{Q^2}
\Bigg) \Bigg]  
\nonumber \\
&&+\frac{\alpha_s}{4\pi}C_F\int\limits_0^{s_0}\!\! \frac{ds\,s  e^{-s/{M^2}}}{Q^4} 
\left\{
(2\ln\frac{s}{\mu^2}-3)\int\limits_0^1 du 
 \left[ \frac{\varphi_{\pi}(u)-\bar u\,\varphi'(0)}{\bar u^2}\right]\right.
\nonumber \\
&&{}\hspace*{4cm}+\left. (2 \ln\frac{s}{\mu^2}-8) 
\int\limits_0^1\!\! du  \frac{\varphi_{\pi}(u)}{\bar u}\right\}.
\label{3:SRexpand}
\end{eqnarray}
The scale dependence cancels to the required accuracy, since \cite{BBB98}  
\begin{equation}
\frac{d}{d\ln\mu} \varphi_\pi'(0,\mu) = \frac{\alpha_s}{\pi}
\Bigg\{ \int\limits_0^1du \left[
\frac{\varphi_\pi(u) + \bar u \varphi_\pi'(1)}{\bar u ^2} 
+ \frac{\varphi_\pi(u)}{\bar u} \right] - \frac{1}{2}\varphi_\pi'(1)\Bigg\}
\;.
\end{equation}
The contribution of hard rescattering equals 
\begin{eqnarray}
F_{\pi}^{\rm hard} (Q^2) & = &
\frac{\alpha_s}{2\pi}C_F\int\limits_0^{s_0} \frac{ds \, e^{-s/{M^2}}}{Q^2} 
\int\limits_0^1 du \frac{\varphi_{\pi}(u)}{\bar u}
\nonumber\\
&&{}+\frac{\alpha_s}{4\pi}C_F\int\limits_0^{s_0}\!\! \frac{ds\,s\,
  e^{-s/{M^2}}}{Q^4} 
\left\{
(2\ln\frac{s}{\mu^2}-3)\int\limits_0^1 du 
 \left[ \frac{\varphi_{\pi}(u)-\bar u\,\varphi'_\pi(0)}{\bar u^2}\right]\right.
\nonumber \\
&&{}\hspace*{4cm}+\left. (2 \ln\frac{s}{\mu^2}-8) 
\int\limits_0^1\!\! du  \frac{\varphi_{\pi}(u)}{\bar u}\right\}
\nonumber\\
&&{}+\frac{\alpha_s}{4\pi}C_F\varphi_{\pi}'(0)
\int\limits_0^{s_0} \frac{ds \, e^{-s/{M^2}}}{Q^4}
 \left\{s\left(2\ln\frac{s}{\mu^2}-3\right)\ln\frac{Q^2}{s_0-s}
-2s_0\right\}\,,
\label{3:Fhard} 
\end{eqnarray}
and the soft contribution is identified as the difference
$F_{\pi}^{\rm soft} (Q^2)= F_{\pi} (Q^2)- F_{\pi}^{\rm hard} (Q^2)$.
Note the term proportional to $\varphi'_\pi(0)= -\varphi_{\pi}'(1)$ 
in the last line of Eq.~(\ref{3:Fhard})  which is concentrated at  
the end-point but enters as  part of the 
hard contribution.

A few comments are in order concerning this expansion.
First, consider the leading asymptotic $O(1/Q^2)$ term. 
 Substituting, as above, $\int_0^{s_0}\!ds\,e^{-s/M^2}\to 4\pi^2f_\pi^2$,
this contribution can be rewritten as 
\begin{equation}
 F_\pi(Q^2) = \frac{8\pi\alpha_s f_\pi^2}{9Q^2}
 \int\limits_0^1 \!\!dv \frac{\varphi^{\rm as}_{\pi}(v)}{\bar v}
  \int\limits_0^1\!\! du \frac{\varphi_{\pi}(u)}{\bar u}\,,
\label{3:nonexact}
\end{equation}   
where we used that 
$\int_0^1 \!\!dv\varphi^{\rm as}_{\pi}(v)/\bar{v} =3$.
This expression is similar, but does not yet coincide with the
perturbative QCD result \cite{exclusive}
\begin{equation}
 F_\pi(Q^2) = \frac{8\pi\alpha_s f_\pi^2}{9Q^2}
  \left|\int\limits_0^1\!\! du \frac{\varphi_{\pi}(u)}{\bar u}\right|^2\,.
\label{3:exact}
\end{equation}
It is easy to convince oneself that the missing corrections to the 
asymptotic pion distribution amplitude in the first integral in 
Eq.~(\ref{3:nonexact}) as well as the missing nonperturbative corrections 
to $f_\pi$ are supplied by eventual higher-order and higher-twist
corrections to the sum rule. Since such corrections are difficult to 
evaluate directly, one may try to improve the light-cone sum rule 
by combining it with the known full NLO perturbative calculation.
Such a possibility will be discussed in Section~6.
 
Second, the structure of the $O(1/Q^4)$ power correction to the pion form 
factor is very similar to the heavy quark limit of the light-cone 
sum rule for $B\to\pi e\nu$ decay considered in \cite{BBB98}.
In particular, note the $1/\bar u^2$ weight factor in the integral 
over the pion distribution amplitude, the structure of 
double-logarithms  and, finally, the  cancellation of the  collinear
factorization scale-dependence by the same mechanism.
In both cases, the distinction between soft and hard contributions 
necessitates a kind of  generalized ``plus-distribution'' subtraction
of divergent integrals over the pion distribution amplitude
at $u\to 1$, as it is done  
in the second line of Eq.~(\ref{3:Fhard}). This feature seems to 
be general, whereas the distribution of finite terms 
$\sim \varphi'_\pi(0)$ between 
the hard and the soft contributions is arbitrary. The expression for such 
terms  in the last line in Eq.~(\ref{3:Fhard}) corresponds to the 
particular definition (\ref{3:NLO}) with a rigid cutoff in momentum 
fraction. Although such a definition is the most intuitive one, 
it is not unique and, as seen from Eqs.~(\ref{3:SRexpand}), 
(\ref{3:Fhard}), introduces 
rather cumbersome ``surface terms'' $\sim \varphi'_\pi(0)$ which appear
both in hard and soft contributions and cancel in their sum.    
An interesting alternative \cite{BBB98} which we do not pursue in detail
in this work is to define the separation
between hard and soft contributions order by order in the 
$1/Q^2$ expansion using ``plus-distributions'' 
to regularize the divergent momentum-fraction integrals. To the accuracy
of Eq.~(\ref{3:SRexpand}), this procedure corresponds to the definition
of the hard contribution as given in the first three lines in 
Eq.~(\ref{3:Fhard}) omitting the ``surface term''. The soft contribution
is given then by the second line in  Eq.~(\ref{3:SRexpand}).

%%%%%%%%%%%%%%%%%%%% FIGURE 4 SOFT-HARD SEPARATION %%%%%%%%%%%%%%%%%%%%%%
\begin{figure}[tb]
\vspace{1cm}
\centerline{\psfig{figure=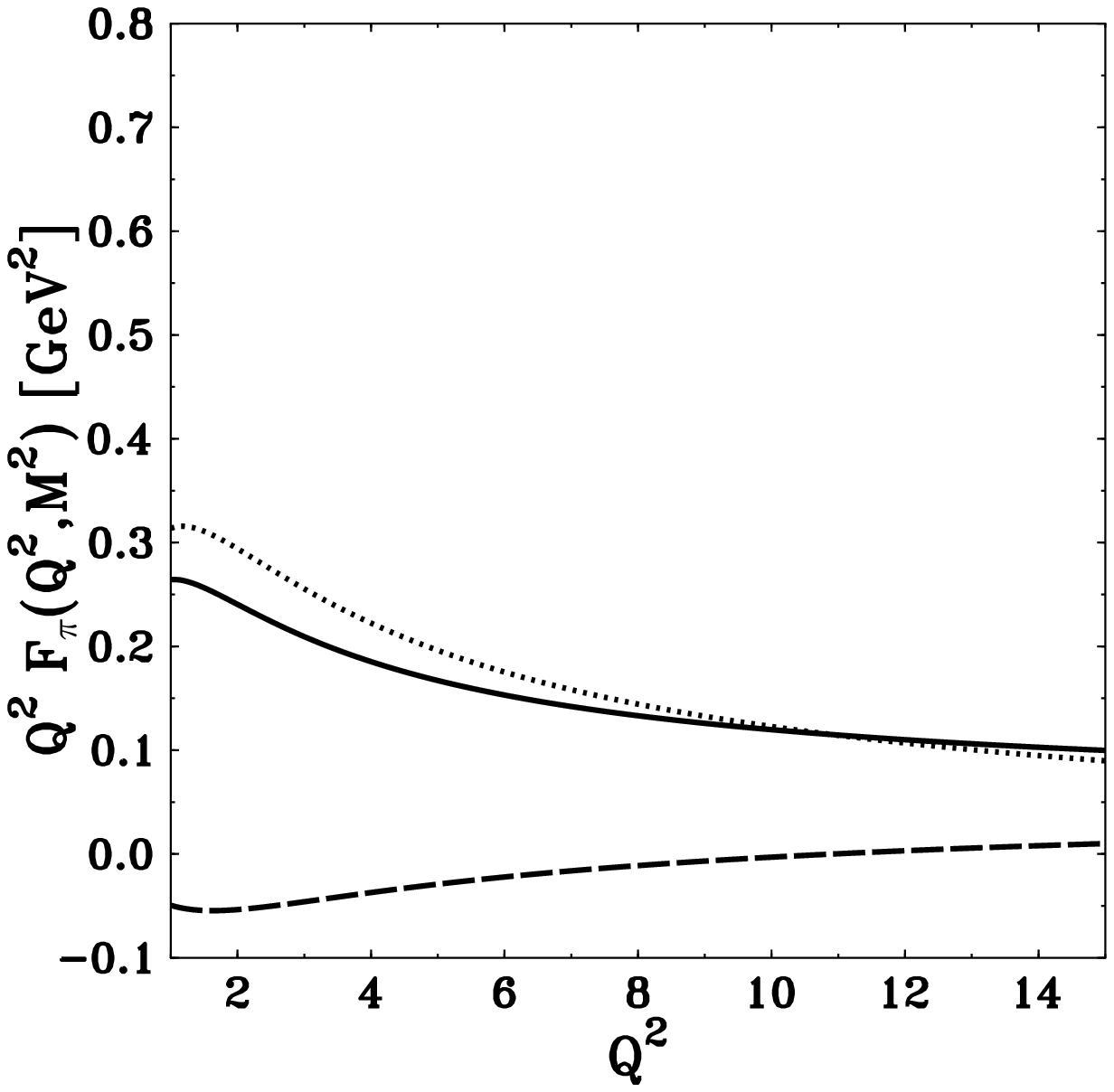,width=6cm}\hspace*{-0.8cm}
            \psfig{figure=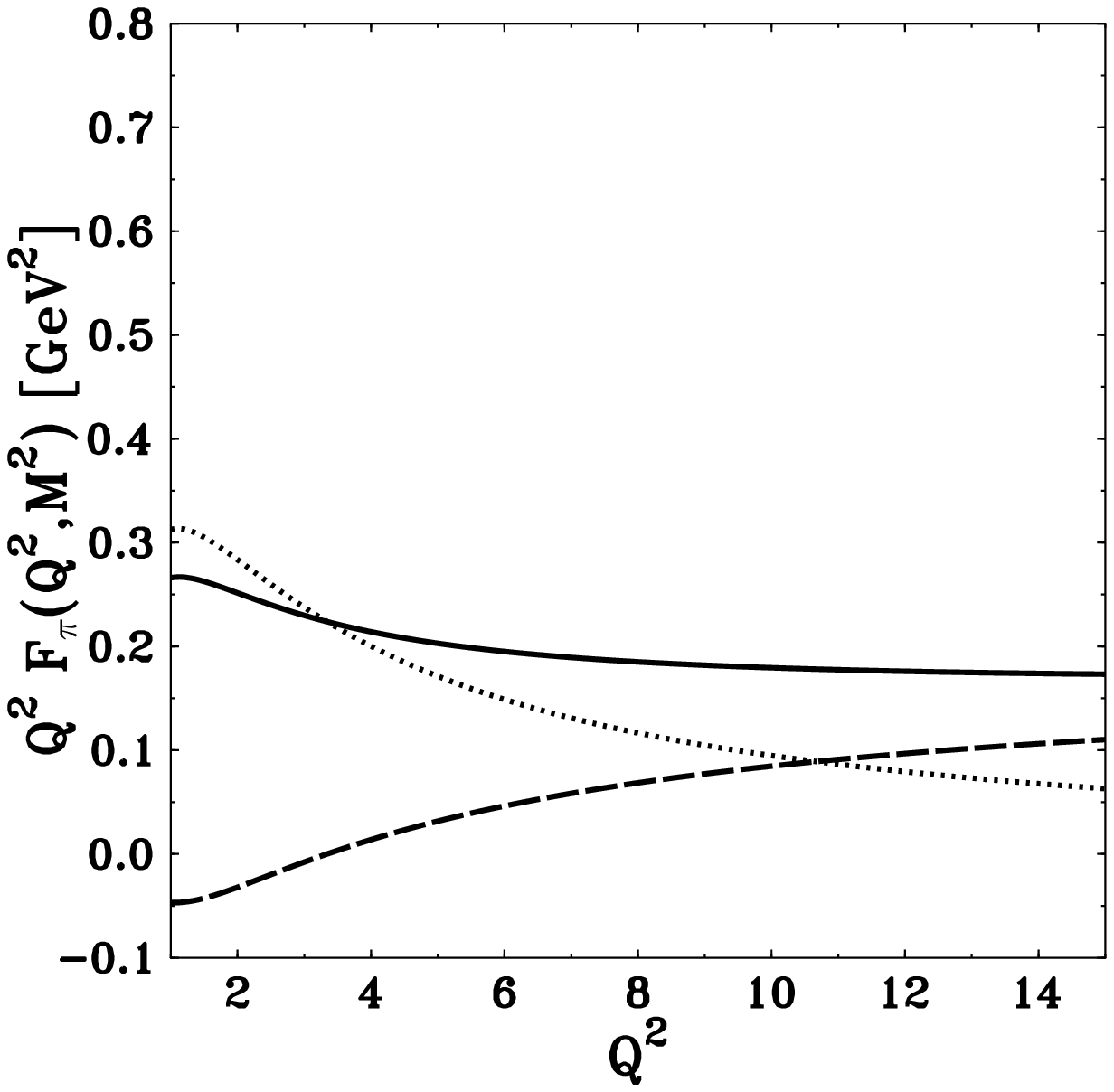,width=6cm}\hspace*{-0.8cm}
            \psfig{figure=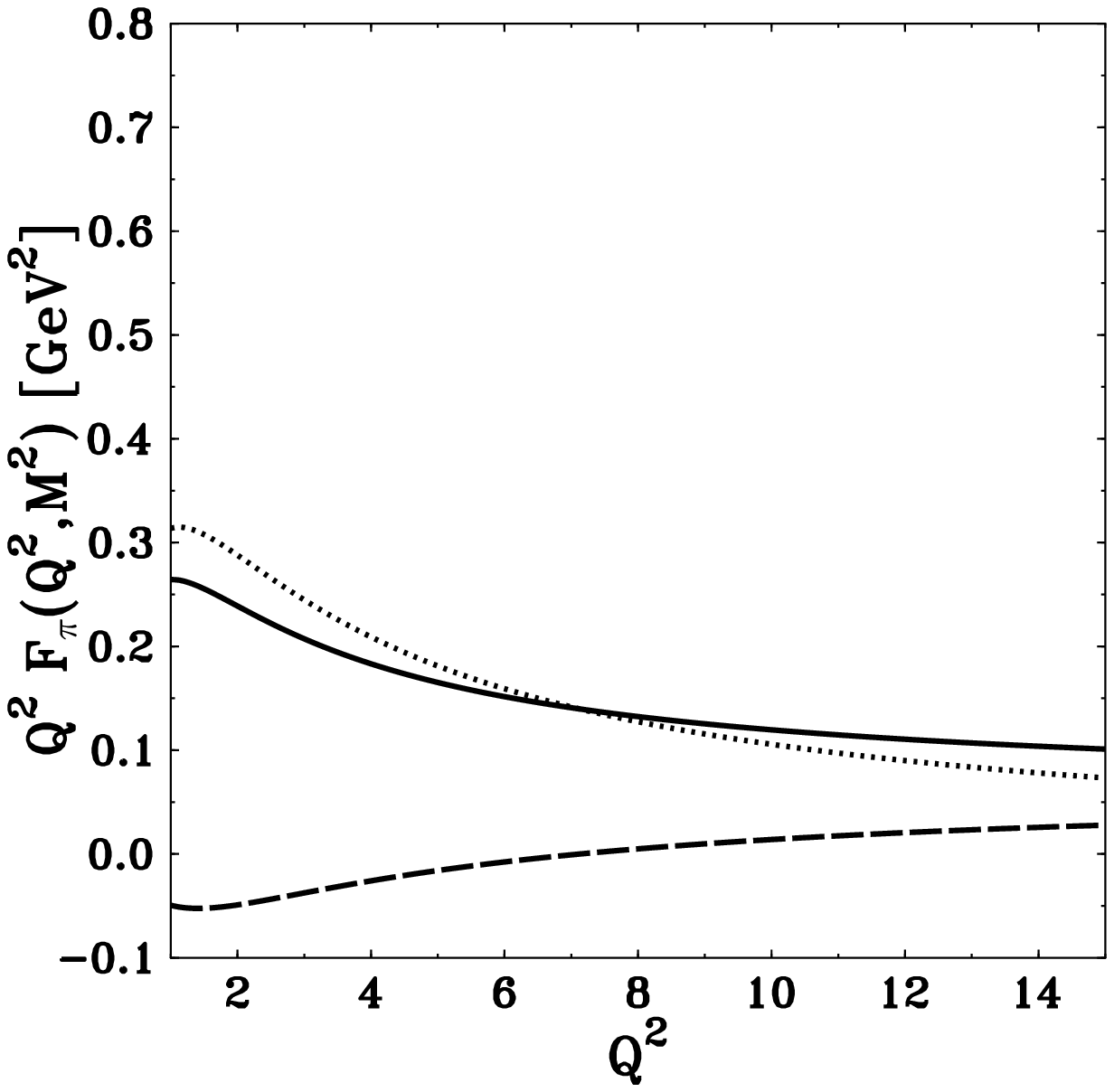,width=6cm}}
\centerline{\psfig{figure=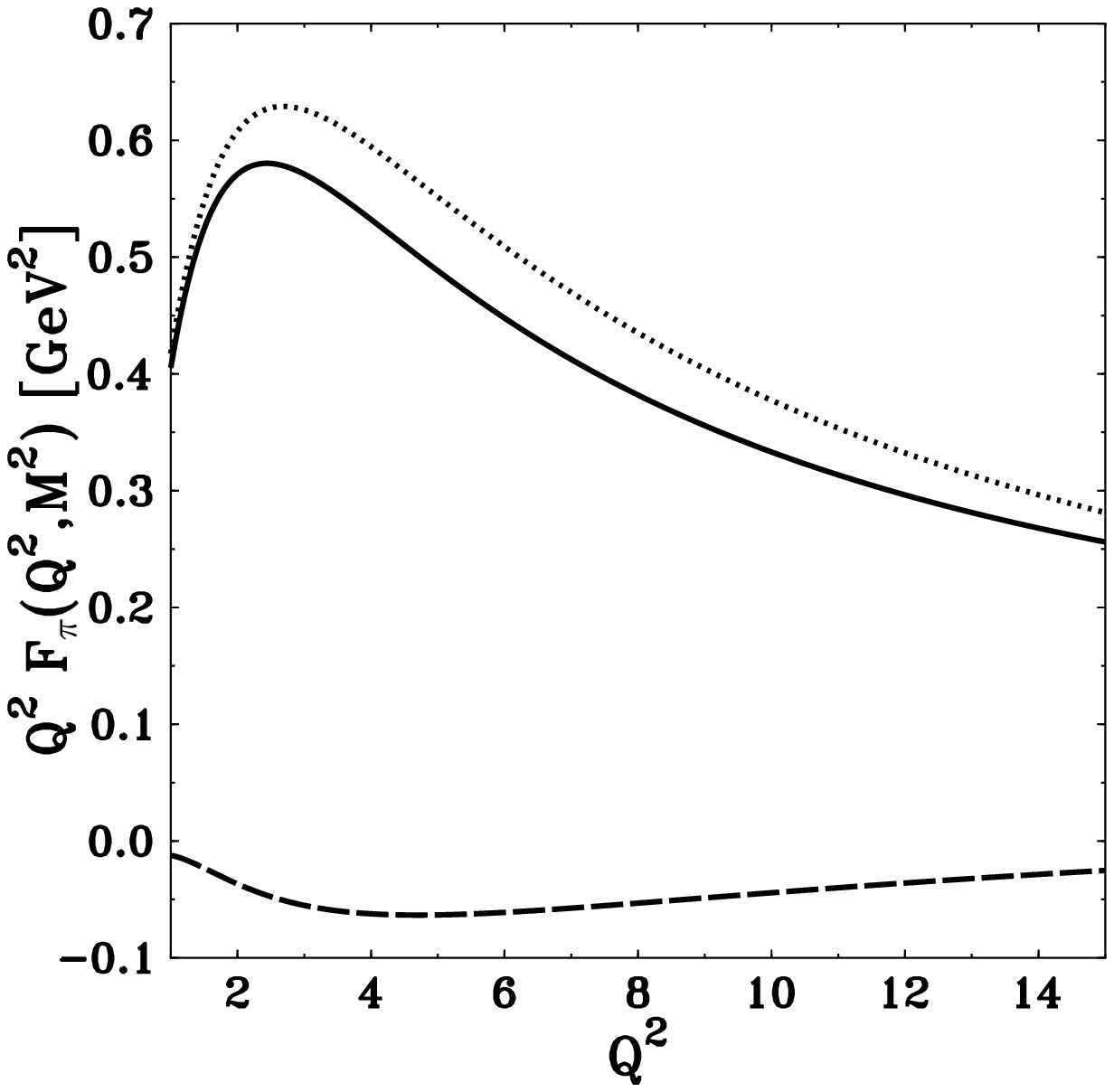,width=6cm}\hspace*{-0.8cm}
            \psfig{figure=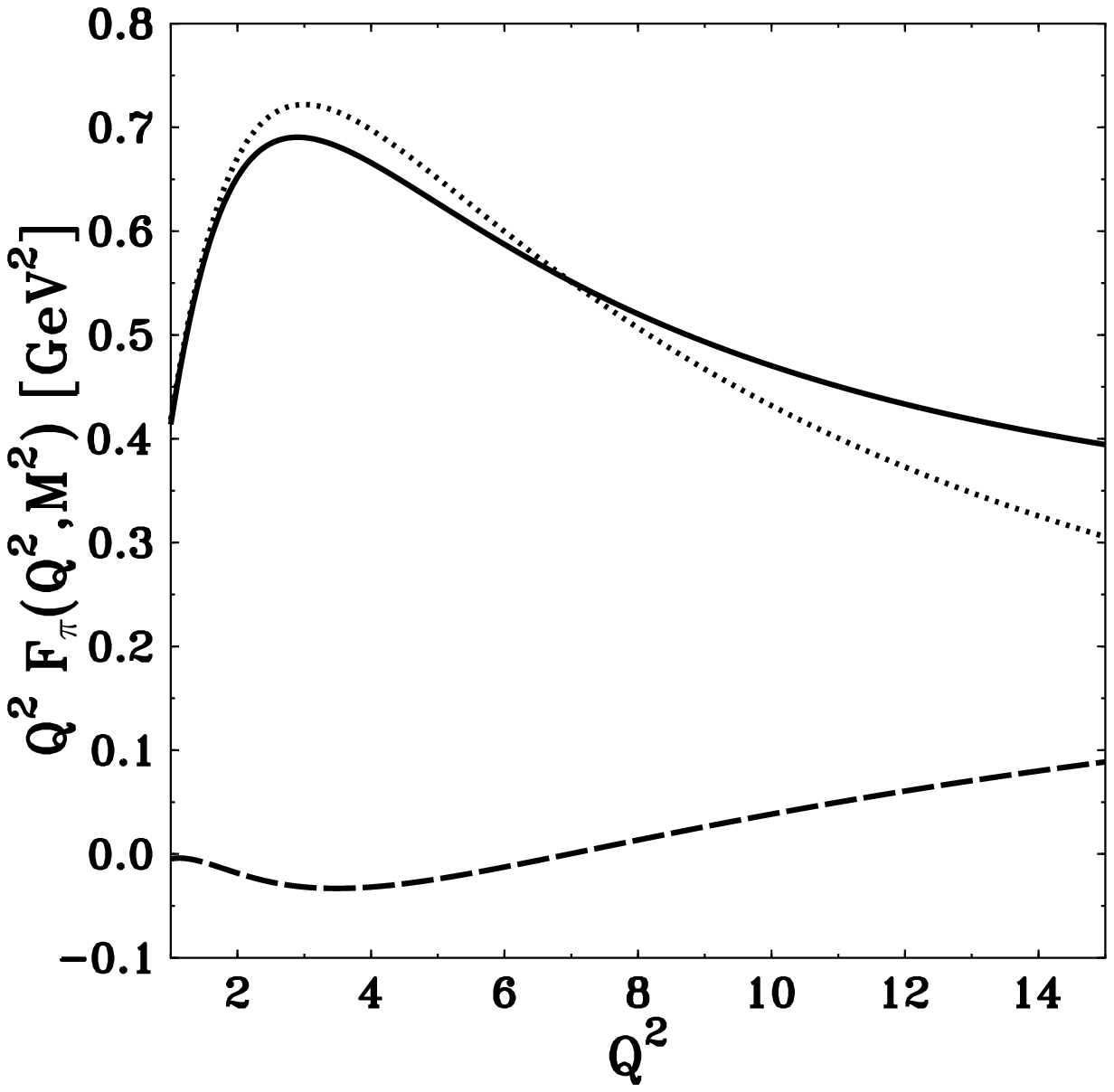,width=6cm}\hspace*{-0.8cm}
            \psfig{figure=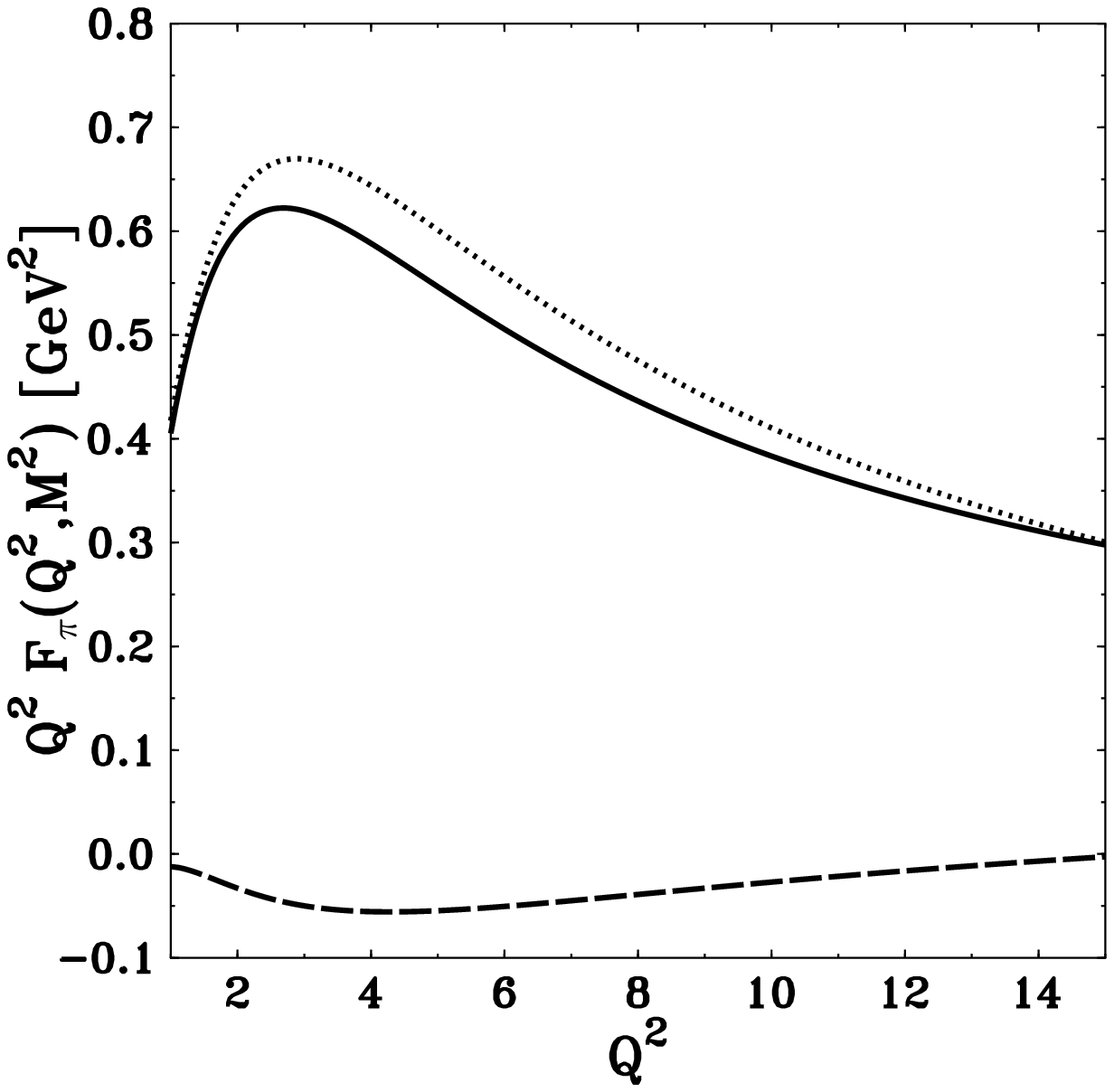,width=6cm}}
\caption
{\small The relative contributions of hard (dashed) and soft (dotted) 
 regions in the leading-twist light-cone sum rule (\protect{\ref{3:NLO}})
 for the pion form factor (the sum: solid lines). The upper figures 
  correspond to asymptotic and the lower figures to
  CZ pion distribution amplitude.
  The three figures from left to right correspond to the choice of 
  factorization scale  $\mu^2 = Q^2$, $\mu^2= s_0$ and 
  $\mu^2= \bar u Q^2 + u M^2$, respectively.
 } 
\label{pertas10}
\end{figure}
%%%%%%%%%%%%%%%%%%%% END FIGURE 4 SOFT-HARD SEPARATION %%%%%%%%%%%%%%%%%%%%%%
%
%

Last but not least, having in mind that the 
separation of hard and soft contributions
is ambiguous, one may add them together and  consider their sum 
as a `total nonperturbative'
power correction to each order in the $1/Q^2$ expansion.
Inspection of Eqs.~(\ref{3:expand}), (\ref{3:localhard}), 
(\ref{3:localsoft}) suggests that soft and hard corrections in general 
have opposite signs and partially cancel in the sum.
We postpone the detailed discussion of this issue to Sect.~6 where we 
summarize our numerical results.

\subsection{Numerical estimates}

Results of the numerical evaluation of the sum rule (\ref{3:NLO})   
are shown in Fig.~4 by solid curves, for $s_0=0.7$~GeV$^2$ and for 
a typical value of the Borel parameter $M^2=1.0$~GeV$^2$. 
(The choice of input parameters is discussed below in Sect. 5).
Soft and hard contributions are  shown by
dotted and dashed curves, respectively.
The results are plotted 
using the asymptotic $\varphi_\pi^{\rm as}(u) = 6u(1-u)$
and the Chernyak-Zhitnitsky (CZ) 
$\varphi_\pi^{\rm CZ}(u,\mu= 1$\,GeV$) = 30\,\bar u u\,(2u-1)^2$ 
pion distribution  amplitudes, for three different choices of the 
factorization scale: $\mu^2=Q^2$, $\mu^2 = s_0$ (see the discussion above) and 
$\mu^2 = \mu^2_u = \bar u Q^2+ u M^2$, according to Eq.~(\ref{2:scale})
\footnote{ If the momentum-fraction dependent scale $\mu_u$ is used, 
it is implied that $\alpha_s(\mu_u)$ is inserted  inside the 
$u$-integrals. }.
In all calculations in this paper we use the two-loop QCD running coupling
with $\Lambda_{\overline{\rm MS}}^{(3)}= 336$~MeV corresponding to 
$\alpha_s(1$\,GeV) = 0.48 and  $\alpha_s(s_0) = 0.59$.   
%
%
%%%%%%%%%%%%%%%%%%%% FIGURE 5 average u  %%%%%%%%%%%%%%%%%%%%%%
\begin{figure}[tb]
\vspace{1cm}
\centerline{\psfig{figure=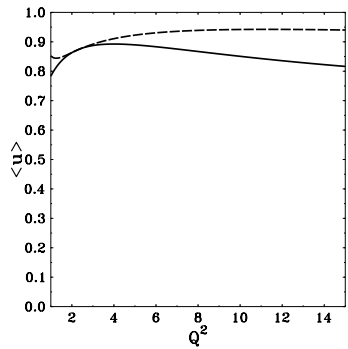,width=6cm}\hspace*{-0.8cm}}
\caption
{\small Average momentum fraction $u$ as a function of $Q^2$ for the 
asymptotic (solid) and CZ (dashed) distribution amplitudes at 
$M^2= 1.0$ GeV$^2$. The scale is $\mu^2= (1-u)Q^2 + uM^2.$ }
\end{figure}
%%%%%%%%%%%%%%%%%%%% END FIGURE 5 average u  %%%%%%%%%%%%%%%%%%%%%%
%
%
It is seen that hard contribution to the form factor defined with a
``natural'' momentum-fraction cutoff remains small and {\em negative}
for the main part of the interesting region of $Q^2$. 

Furthermore, in Fig.~5 we show the average value of the momentum fraction 
$u$ in the integral in Eq.~(\ref{3:NLO}) calculated as a function of $Q^2$ 
for the asymptotic (solid curve) and CZ (dashed curve) 
distribution amplitudes.  This  average value 
turns out to be very large, and, contrary to usual 
expectations, does not depend significantly on the shape of the pion 
distribution amplitude. 

Negative contribution of the hard-rescattering mechanism may appear 
unexpected and counterintuitive. We emphasize, however, that the 
separation between hard and soft terms is ambiguous and depends on their 
definition --- this is, in fact, 
the main lesson to be learnt from our analysis. 
Note that the scale dependence is much more pronounced 
for hard and soft contributions taken separately 
than for their sum. 

\section{Higher-twist corrections}

\subsection{Twist 4 contributions}

The operator-product expansion of the correlation 
function (\ref{2:cor}) near the 
light-cone $x^2=0$ can be continued beyond the 
leading twist 2 approximation (\ref{2:ex6}). This procedure 
yields higher-twist corrections  to the 
light-cone sum rule (\ref{3:NLO}). They are suppressed 
by additional inverse powers of $M^2$ and $Q^2$.
Physically, the higher-order terms of the light-cone expansion
take into account both the transverse momentum of the  
quark-antiquark state and
the contributions of higher Fock states in the pion wave function. 
As explained in \cite{BF90}, these two effects are indistinguishable
due to QCD equations of motion .    
  
Next to the leading twist 2 term, the correlation function 
(\ref{2:cor}) receives  several   twist 4 contributions 
\footnote{Twist 3 contributions to (\ref{2:cor}) are
proportional to $m_\pi^2$ and vanish in the chiral limit adopted here.}.
First of all, one has to take into account the twist 4 components of 
the quark-antiquark matrix element 
$\langle 0| \bar d(0) \gamma_{\mu} \gamma_5 u(x) | 
\pi^+(p) \rangle $ in the diagram in Fig.~1. 
Furthermore, the gluon emission from the virtual quark 
should be included yielding the diagram of Fig.~6a with the twist 4 
quark-antiquark-gluon distribution amplitudes of the pion.
To calculate this diagram, one makes use of the  
light-cone expansion of the quark propagator \cite{BB8889}
given in Appendix A. 
The definitions of all relevant twist 4  two- and three-particle 
distribution amplitudes \cite{BF90,Ball98} are collected in Appendix B. 

The corresponding calculation has been carried out in \cite{BH94}.
The twist 4 contribution  to the correlation function (\ref{2:cor}) 
can be written in the following compact form  
\begin{equation}
T^{(4)}_{\mu\nu}=2ip_{\mu }p_{\nu} f_\pi\int\limits_0^1 du
\frac{u\, \varphi^{(4)}(u)}{\left(\bar{u}Q^2-u s \right)^2}\,,
\label{4:tw4corr}
\end{equation}
where 
\begin{eqnarray}
\varphi^{(4)}(u) &=& -4\left(g_1(u) - \int\limits_0^u dv \,g_2(v) \right)
+2u\,g_2(u) 
+\int \limits_{0}^{u}d\alpha_1 \int\limits_0^{\bar{u}}
d\alpha_2 \Bigg[\frac{\tilde{\varphi}_{\parallel}(\alpha_i)+
2\tilde\varphi_{\perp}(\alpha_i)}{\alpha_3}
\nonumber
\\
&&
+\frac{1-2u+\alpha_1-\alpha_2}{\alpha_3^2}
\left( {\varphi}_{\parallel}(\alpha_i)+ \varphi_\perp(\alpha_i)
\right)\Bigg]_{\alpha_3=1-\alpha_1-\alpha_2} 
\label{4:phi4}
\end{eqnarray}
is a combination of twist 4  
distribution amplitudes of the pion. The explicit expression 
for $\varphi^{(4)}$ is given in  Appendix B. 

%
%%%%%%%%%%%%%%%%%%%% FIGURE 6 propagator expansion DIAGRAM %%%%%%%%%%%%%%%%%%%%%%%%%%
\begin{figure}
\centerline{\epsfig{file=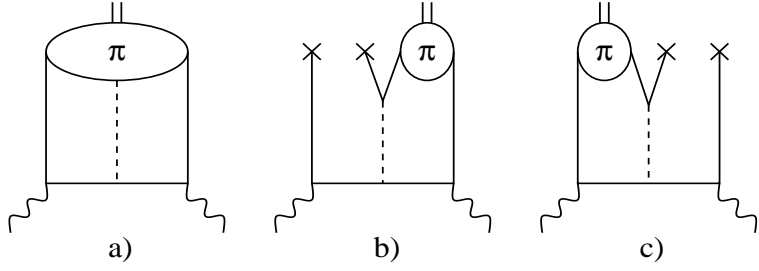, height=3.5cm, width=10.cm}}
\caption[]{\small
 The light-cone expansion of the quark propagator in the correlation function  
 (\ref{2:cor}).}
\end{figure}
%%%%%%%%%%%%%%%%%%%% END OF FIGURE 6 %%%%%%%%%%%%%%%%%%%%%%%%%%
%

The twist 4 correction to the light-cone sum rule is easily 
obtained by taking the imaginary part of Eq.~(\ref{4:tw4corr}) 
in $s=(p-q)^2$ and subtracting the continuum above $s_0$. After Borel
transformation one obtains: 
\begin{equation}
F_\pi^{(4)}(Q^2)= \int^1_{u_0}du \frac{\varphi^{(4)}(u)}{uM^2}
\exp\left( -\frac{\bar{u}Q^2}{uM^2}\right) + \frac{u_0\varphi^{(4)}(u_0)
}{Q^2}e^{-s_0/M^2}\,.
\label{4:tw4res}
\end{equation}
The second term in the r.h.s. of Eq.~(\ref{4:tw4res}) has not been taken
into account in \cite{BH94}. It appears as a `surface term'
 when the correlation function with a denominator 
$(q-up)^{2n}= (-\bar{u}Q^2+us)^n$ with $n>1$  
is converted (integrating by parts) into a canonical 
dispersion integral $T_{\mu\nu}(s,Q^2) = 1/\pi\int d\tilde{s} 
\,\mbox{Im} T_{\mu \nu}(\tilde{s},Q^2)/(\tilde{s}-s)$.    
Adding the expression in Eq.~(\ref{4:tw4res}) to the leading twist 2 
contribution  (\ref{3:NLO})
one obtains the light-cone sum rule for $F_\pi(Q^2)$ to the twist 4 accuracy.
As seen from Eq.~(\ref{4:tw4res}), all twist 4 effects have to be identified 
(to our accuracy) as part of the soft contribution to the form factor.
Since $\varphi^{(4)}(u)\sim (1-u)$ at $u\to1$ (see Appendix B), 
the twist 4 corrections are of order $1/Q^4$ in the large $Q^2$ limit.  

Assuming asymptotic expressions for the quark-antiquark-gluon distribution
amplitudes \cite{BF90,Ball98,CZ84} one obtains a compact expression:
\begin{equation}
\varphi^{(4)}(u)= \frac {20}3\delta^2(\mu) u^2\,\bar{u}\,(3u-2)\,,
\label{tw4asy}
\end{equation}
where $\delta^2(1 \mbox{GeV})\simeq 0.2$~GeV$^2$ 
is a scale-dependent parameter 
determining the pion coupling  
to the local quark-antiquark-gluon operator (see Appendix B 
for the definition). The twist 4 correction to the sum rule  
simplifies in  this case to: 
\begin{equation}
F_\pi^{(4)}(Q^2)= \frac {20}3\delta^2(\mu)\int\limits_0^{s_0}ds\,e^{-s/M^2}
\frac{Q^8}{(Q^2+s)^6}\left( 1-\frac{8s}{Q^2}+ \frac{6s^2}{Q^4} \right),
\label{tw4asy2}
\end{equation}
revealing at $ Q^2 \to \infty$  the 
$1/Q^4$ behavior.
Taking in addition the local duality limit $ M^2 \to \infty$ yields
an estimate
\begin{equation} 
F_\pi^{(4)}(Q^2)=\frac {20\delta^2(s_0)s_0}{3Q^4} \sim 
    \left(\frac{1.0\, \mbox{GeV}^2}{Q^2}\right)^2.
\label{tw4asy2ld}
\end{equation}

\subsection{Factorizable twist 6 contributions}

An estimate of twist 6 contribution to the light-cone 
sum rule presents a new result of this paper.
This calculation is interesting for several reasons.
As well known \cite{SV82}, twist 4 operators are `irreducible' in the sense 
that they cannot be factorized in a product of gauge-invariant operators of
lower twist. This property is special and limited to twist 4.
 Several light-cone operators of twist 6 exist which 
can be factorized as a product of two gauge-invariant twist 3 operators 
(or, alternatively,  one twist 2 and one twist 4). Sandwiched between 
vacuum and one-pion state, such operators generally produce 
two types of contributions: 
Factorizable in terms of a low-twist two-particle 
distribution amplitude times quark (or gluon)  condensate, and
nonfactorizable that give rise to genuine 
twist 6 multiparton pion distribution amplitudes.
We emphasize that factorizable contributions have to be subtracted 
in the construction of multiparton distribution amplitudes similar 
as disconnected diagrams proportional to the quark condensate 
should not be taken into account in the nucleon matrix element 
 $\langle N| \bar q q |N\rangle$ corresponding to the nucleon $\sigma$-term. 

In the present context, arguments based on conformal symmetry suggest 
that contributions of  higher Fock states are strongly suppressed at 
$u\to 1$ and their contributions to the sum rule are, probably, 
negligible. Factorizable contributions, on the other hand, are expected to
supply the missing nonperturbative corrections in the sum rule 
in the large $Q^2$ limit and can be large. They are also of principal 
interest and indicate, as we will see, certain limitations for the light-cone 
sum rule approach. 
%
%
%%%%%%%%%%%%%%%%%%%% FIGURE 7 cat ear DIAGRAM %%%%%%%%%%%%%%%%%%%%%%%%%%
\begin{figure}
\centerline{\epsfig{file=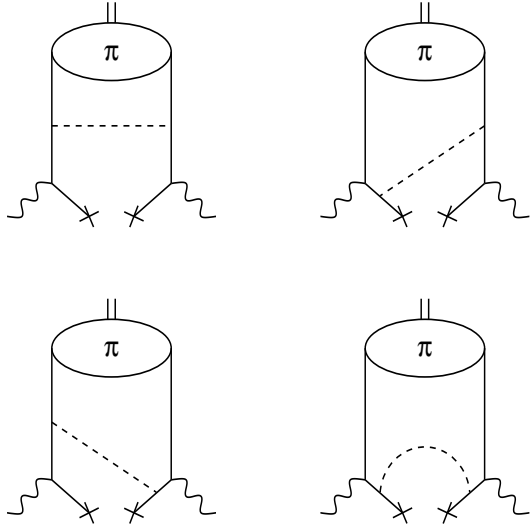, height=7.0cm, width=7.cm}}
\caption[]{\small
 Quark condensate corrections to the correlation function  
 in Eq.~(\protect{\ref{2:cor}}).
   }
\end{figure}
%%%%%%%%%%%%%%%%%%%% END OF FIGURE 7 %%%%%%%%%%%%%%%%%%%%%%%%%%

Guided by existence of large quark condensate corrections 
$\sim \langle 0|\bar q q|0\rangle^2$ in classical QCD sum rule calculations
of the pion form factor \cite{IofSmil,Rad82}, 
in this paper we concentrate on factorizable 
contributions of twist 6 four-quark operators, e.g.
\begin{eqnarray}
\lefteqn{\hspace*{-2cm}
\langle 0| \bar q(v_1x)\gamma_\mu {q}(v_2 x) \bar q(v_3 x)\gamma_\nu 
\gamma_5q(v_4 x) 
| \pi^+(p) \rangle =}
\nonumber\\
&=&\frac{i}{12} \langle 0|\bar{q}q |0\rangle
\langle 0|\Big[\bar q(v_1 x)\sigma_{\mu\nu}\gamma_5 q(v_4x)+
               \bar q(v_3 x)\sigma_{\mu\nu}\gamma_5 q(v_2x)\Big]
| \pi^+(p) \rangle 
\nonumber\\
&&{}-\frac{1}{12} g_{\mu\nu} \langle 0|\bar{q}q |0\rangle
\langle 0|\Big[\bar q(v_1 x)\gamma_5 q(v_4x) -
               \bar q(v_3 x)\gamma_5 q(v_2x)\Big]
| \pi^+(p) \rangle\,, 
\label{4:4quark}
\end{eqnarray} 
which involve the quark condensate and the two existing two-particle 
pion distribution amplitude of twist 3, $\varphi_p$ and $\varphi_\sigma$,
see Appendix B. Definitions of the both of them 
include the normalization factor   
\begin{equation}
\mu_\pi=\frac{m_\pi^2}{m_u+m_d}=-\frac{2}{f_\pi^2}\langle\bar q q \rangle\,,
\label{mupi}
\end{equation}
so that the corresponding contributions to the sum rule appear to be 
 proportional to the quark condensate squared.   

We start from the light-cone expansion of the quark propagator
(see Appendix A) which contains contributions proportional to 
the covariant derivative $D^\alpha G_{\alpha\nu}$ of the gluon field strength.
They are reduced to a quark-antiquark
pair due to the QCD equations of motion. One quark (antiquark) 
from this pair can be combined with an antiquark (quark) from 
the initial currents forming a quark condensate as in Fig.~6b,c. 

A straightforward calculation gives:
\begin{eqnarray}
T_{\mu\nu}^{\rm Fig.6b,c} &=& 2i p_{\mu}p_{\nu}  \alpha_s \pi C_F
\frac{\langle \bar q q\rangle}{N_c} f_\pi \mu_\pi \int\limits_0^1\!\!du 
\!\int\limits_0^1 \!\!\!dv \,v\bar v 
 \Bigg(2 \varphi_p(u) \Bigg\{\frac{(1\!-\!uv)(u\!-\!2)}{[q-(1\!-\!uv)p]^6}
                                 +\frac{vu^2}{[q-uvp]^6}\Bigg\}
\nonumber \\
&&\hspace{-0.8cm}{}+ 
 \frac{1}{3}\varphi_\sigma(u)\Bigg\{\frac{uv-3}{[q-(1\!-\!uv)p]^6}
                                - \frac{uv}{[q-uvp]^6} 
                + \frac{3q^2 v(2+u)}{[q-(1\!-\!uv)p]^8}
                - \frac{3q^2 uv}{[q-uvp]^8}\Bigg\} \Bigg).
\label{propexp}
\end{eqnarray}
Another source of the factorizable twist~6 contribution is provided 
by the four-quark operators in the light-cone expansion 
of Eq.~(\ref{2:cor}) with a perturbative gluon exchange between two currents 
(see Fig.~7).
The technique of this expansion is explained in \cite{BB8889,Ba83}. 
A lengthy but equally straightforward calculation yields: 
\begin{eqnarray}
T_{\mu\nu}^{\rm Fig.7 } &=& 8ip_{\mu}p_{\nu} \alpha_s \pi C_F 
\frac{\langle \bar q q\rangle}{N_c} \mu_{\pi} f_{\pi} 
\int\limits_0^1\!\! du\,\int\limits_0^1\!\!\! dv\,v\!
\Bigg[ \varphi_p(u)\frac{\bar u}{[q-(1\!-\!uv)p]^6} 
\nonumber \\
&& {} +\frac{1}{2}\varphi_\sigma(u)\,
               \frac{1}{[q-(1\!-\!uv)p]^6}   
    +\varphi_\sigma(u)\,\frac{v(\bar u pq -q^2)}{[q-(1\!-\!uv)p]^8}
 \Bigg].
\label{fig6}
\end{eqnarray}
In addition, we have considered the twist 6 parts of the two- and 
three-particle matrix elements corresponding to  
the diagrams of Fig.~1 and Fig.~5a and have not found any
factorizable contributions.
The sum of Eqs.~(\ref{propexp}) and (\ref{fig6}) represents, therefore, 
the complete answer for the factorizable twist 6 contributions
of four-quark operators.

The corresponding correction to the light-cone sum rule 
can be obtained following the standard procedure, that is
taking imaginary part in $s=(p-q)^2$, 
subtracting the continuum above $s=s_0$   
in the dispersion integral, and performing the Borel transformation.
Due to large dimension of the denominators in Eqs.~(\ref{propexp}) and 
(\ref{fig6}), 
one ends up with a rather complicated structure of surface terms at 
$s=s_0$. The final answer can be written as 
\begin{eqnarray}
F^{(6)}_\pi(Q^2) &=& \frac{\alpha_s \pi C_F}{N_c} 
\langle\bar{q} q\rangle\,\mu_{\pi}\!\int\limits_0^{\infty}\! 
ds \Bigg[ f_2(s,Q^2)\frac{d^2}{ds^2}
\left( \Theta(s_0-s)e^{-s/M^2}\right)
\nonumber
\\
&&
+f_3(s,Q^2)\frac{d^3}{ds^3}
\left( \Theta(s_0-s)e^{-s/M^2}\right) \Bigg],
\label{prop1}
\end{eqnarray}
where 
\begin{equation}
f_{2,3}(s,Q^2) = f_{2,3}^{Fig.6bc}(s,Q^2) + f_{2,3}^{Fig.7}(s,Q^2),
\end{equation}
with
\begin{eqnarray}
f_2^{Fig.6bc}(s,Q^2) &=& \frac{s}{2Q^2(Q^2+s)}
  \!\!\!\! \int\limits_{s/(Q^2+s)}^1\!\!\!\!\!
\frac{du}{u^3} \left(u\!-\!\frac{s}{Q^2+s}\right)\!
\left[2(2\!-\!u)\,\varphi_p(u)+
\left( 1+\frac{2s}{3Q^2}\right) \varphi_{\sigma}(u) \right ]    
\nonumber
\\
&&-\frac{1}{2(Q^2+s)}\!\!\!\int\limits_{Q^2/(Q^2+s)}^1\!\!\!
 \frac{du}{u^3}\left(u\!-\!\frac{Q^2}{Q^2+s} \right)
\left[2u\varphi_p(u)-
\frac13\varphi_{\sigma}(u)\right ],
\nonumber
\\
f_2^{Fig.7}(s,Q^2) &=& -\frac{2s}{Q^4}
\!\!\!\int\limits_{s/(Q^2+s)}^1\!\!\!
\frac{du}{u^2}\left[ \bar{u}\varphi_p(u) + \frac12 \varphi_{\sigma}(u)
\left ( 1 - \frac{\bar{u}s}{uQ^2}\right) \right],
\label{prop2}
\end{eqnarray}
\begin{eqnarray}
f_3^{Fig.6bc}(s,Q^2) &=&  \frac{s^2}{6Q^4}
\!\!\!\!\int\limits_{s/(Q^2+s)}^1\!\!\!\!\!\!
\frac{du}{u^4}\left(u\!-\!\frac{s}{Q^2+s}\right)(2\!+\!u)\varphi_{\sigma}(u)
-\frac{1}{6}\!
\!\!\!\!\int\limits_{Q^2/(Q^2+s)}^1\!\!\!\!\!\!\!
\frac{du}{u^3}\left(u\!-\!\frac{Q^2}{Q^2+s}\right)\varphi_{\sigma}(u)\,,
\nonumber
\\
f_3^{Fig.7}(s,Q^2) &=& -\frac{2s^2}{3Q^4}\int\limits_{s/(Q^2+s)}^1
\frac{du}{u^3}\,\varphi_{\sigma}(u) \left( 1-\frac{\bar{u}(Q^2+s)}{2Q^2}
\right).
\label{prop3}
\end{eqnarray}
It is easy to see that the expressions in Eqs.~(\ref{prop2}) and 
(\ref{prop3}) receive contributions from both hard $u<u_0$ and soft
$u>u_0$ regions, which we do not write separately in this case.
The hard contribution takes into account the integration region
corresponding to a large momentum $\sim Q$ flowing through the gluon 
line and can be thought of as part of the hard mechanism contribution
to the form factor generated by product of two twist 3 distribution 
amplitudes, with `wrong' quark helicities \cite{GT82,MP94}. 
In the light-cone sum rule 
approach one distribution amplitude is present directly, and the second
one is modelled using the duality approximation, as in the leading twist.

Inserting the asymptotic expressions for the  distribution amplitudes
$\varphi_p(u)= 1$ and $ \varphi_\sigma(u) = 6u (1-u)$  and  integrating 
over $u$ one obtains the expansions 
\begin{eqnarray}
f_2(s,Q^2) &=& -\frac{1}{Q^2} + \frac{s}{Q^4}\left ( 5 -
2\ln\frac{Q^2}{s}\right) + \ldots \,,
\nonumber
\\
f_3(s,Q^2) &=& - \frac{s}{Q^2} + 
\frac{s^2}{Q^4}\left ( 3-\ln \frac{Q^2}{s}\right)+ \ldots\,,
\label{f23asy}
\end{eqnarray}
substitution of which  in Eq.~(\ref{prop1}) 
yields the twist 6 correction to the light-cone sum rule
to the $O(1/Q^4)$ accuracy:
%
%
%\begin{eqnarray}
%F_{\pi}^{(6)}(Q^2) &=& -\frac{\alpha_s \pi C_F}{N_c} 
%\frac{\langle \bar q q \rangle \mu_\pi}{Q^4}\Bigg\{
%1+\Bigg[ \left( 1+\frac{s_0}{M^2} -\frac{s_0^2}{M^4}
%\ln\frac{Q^2}{s_0} \right)e^{-s_0/M^2}
%\nonumber \\
%&& \qquad +\int\limits_0^{s_0} ds e^{-s/M^2} \frac{s}{M^4}
%\left(2-\frac{s}{M^2}\right)\ln \frac{Q^2}{s} \Bigg] \Bigg\}\,.
%\label{tw6fin}
%\end{eqnarray}
%
%%%%%%%
\begin{equation}
 F^{(6)}_\pi(Q^2) = \frac{4\alpha_s \pi C_F}{N_c f_\pi^2 Q^4} 
\langle\bar{q} q\rangle^2~ \simeq
\left(\frac{0.2 \, \mbox{GeV}^2}{Q^2}\right)^2,
\label{eq:t6}
\end{equation}
much smaller than  $F^{(4)}_\pi$.

Most importantly, the $O(1/Q^2)$ contributions  have cancelled. 
Inspection of Eqs.~(\ref{prop2}) and (\ref{prop3}) reveals that all 
$O(1/Q^2)$ contributions in individual diagrams originate from the 'hard' 
integration region $u<u_0$ and their cancellation involves
both diagrams and both distribution functions $\varphi_\sigma$ and 
$\varphi_p$, in agreement with 
\cite{GT82}.

The observed cancellation of $O(1/Q^2)$ corrections is not entirely 
trivial. One might fear that factorization of a local 
quark condensate brings us back to the deficiency of the standard 
QCD sum rule approach discussed in Sect.~2: Expansion of local operators
messes up the power counting in the momentum transfer, as observed in 
\cite{IofSmil,Rad82}. In other language, factorization of the 
quark condensate is equivalent to using a very bad model 
for the distribution amplitude of the pion created by the interpolation 
current in Eq.~(\ref{2:cor}), corresponding to the  sum of 
two $\delta$-functions, see \cite{RR96,BB97}.\footnote{
A detailed comparison with the results of \cite{GT82,MP94} goes beyond
the tasks of this paper. In particular, one may ask whether light-cone 
QCD sum rules can be used to calculate an effective infrared cutoff
in the  hard scattering 
contribution obtained in  \cite{GT82,MP94}. To address this issue one  
has to construct a 
different sum rule, using a chiral-odd interpolating current for the pion.
Interpretation of Eq.~(\ref{eq:t6}) in this context is difficult because
of possible contamination by the $A_1$ meson. 
We thank M. Beneke and G. Buchalla for the discussion which initiated 
our interest to this problem.} 
%
%The remaining $O(1/Q^4)$ contribution presents the sum of soft and hard 
%contributions and is finite. Similar to the leading twist, 
%the separation between hard and soft terms is scale- and scheme-dependent  
%as signalled by  the logarithmic infrared divergence of the hard
%contribution, see \cite{MP94}.

%In the local duality limit $M^2\to\infty$ one obtains
%\begin{equation}
% F^{(6)}_\pi(Q^2) \simeq \frac{4\alpha_s \pi C_F}{N_c f_\pi^2 Q^4} 
%\langle\bar{q} q\rangle^2~ \simeq
%\left(\frac{0.2 \, \mbox{GeV}^2}{Q^2}\right)^2,
%\end{equation}

\section{Numerical analysis}  

Combining the twist~2 calculation (\ref{3:NLO}) with 
the twist~4 corrections in Eq.~(\ref{4:tw4res}) and twist~6  
in Eq.~(\ref{prop1}), we are now in a position to evaluate the complete
light-cone sum rule for the pion form factor:
\begin{equation}
  F_\pi(Q^2) = F_\pi^{(2)}(Q^2,M^2)+F_\pi^{(4)}(Q^2,M^2)
               + F_\pi^{(6)}(Q^2,M^2)\,.
\end{equation} 
To avoid misunderstanding, note that terms of higher twist are 
not suppressed, in general, by increasing powers of $1/Q^2$, but rather
by increasing powers of the Borel parameter. In particular, all 
twists contribute to $1/Q^4$ accuracy, with main contributions 
 coming from the soft region, 
in agreement with the general wisdom (see also Sect.~2) that such 
corrections come from large transverse distances. The (numerical) hierarchy 
of contributions of different twist is, therefore,
a self-consistency check for the light-cone sum rule approach. 
%
%
%%%%%%%%%%%%%%%%%%%%% FIGURE 8  Borel dependence %%%%%%%%%%%%
\begin{figure}[tb]
\vspace{-1cm}
\centerline{\psfig{figure=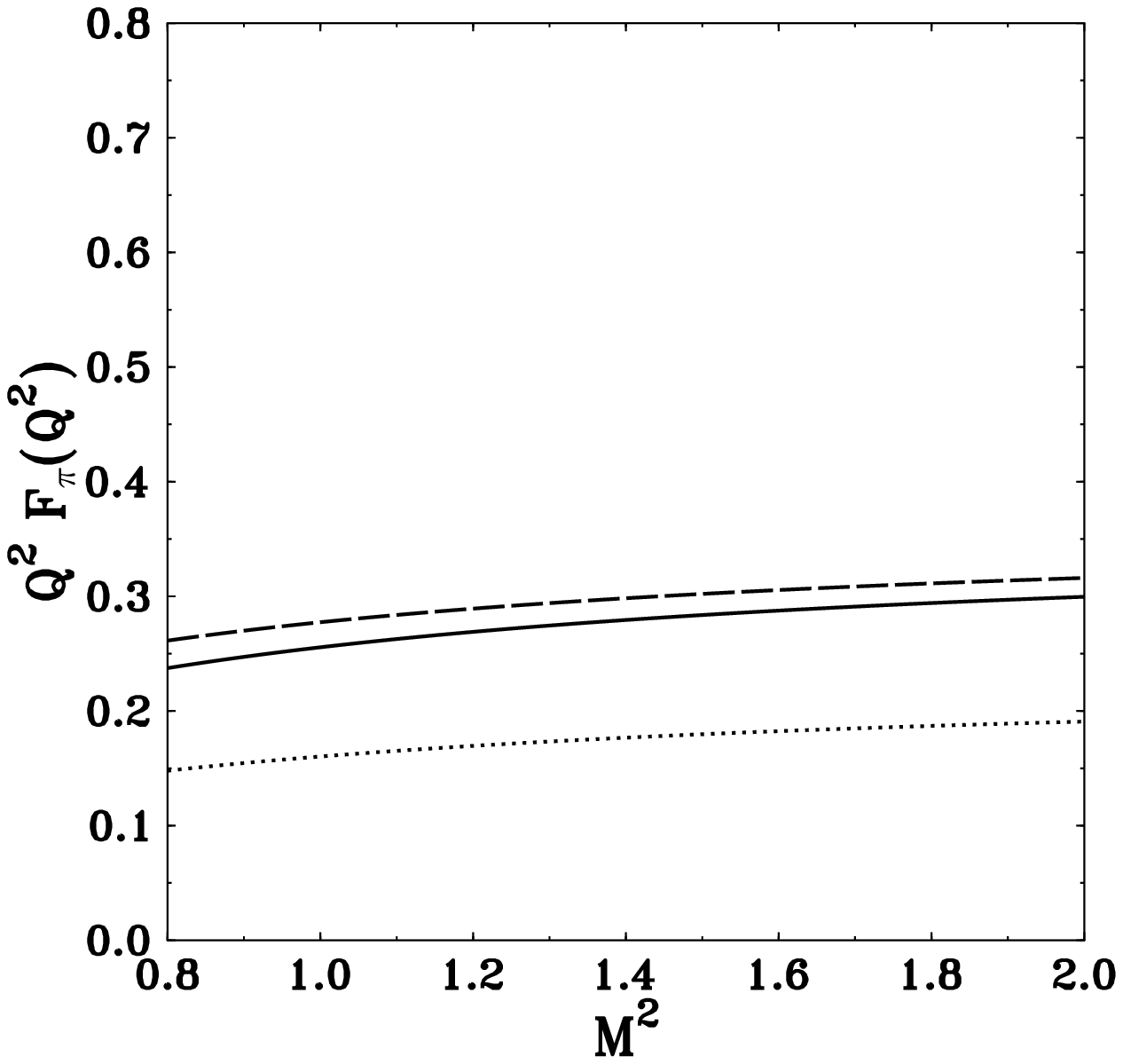,width=8cm}
            \psfig{figure=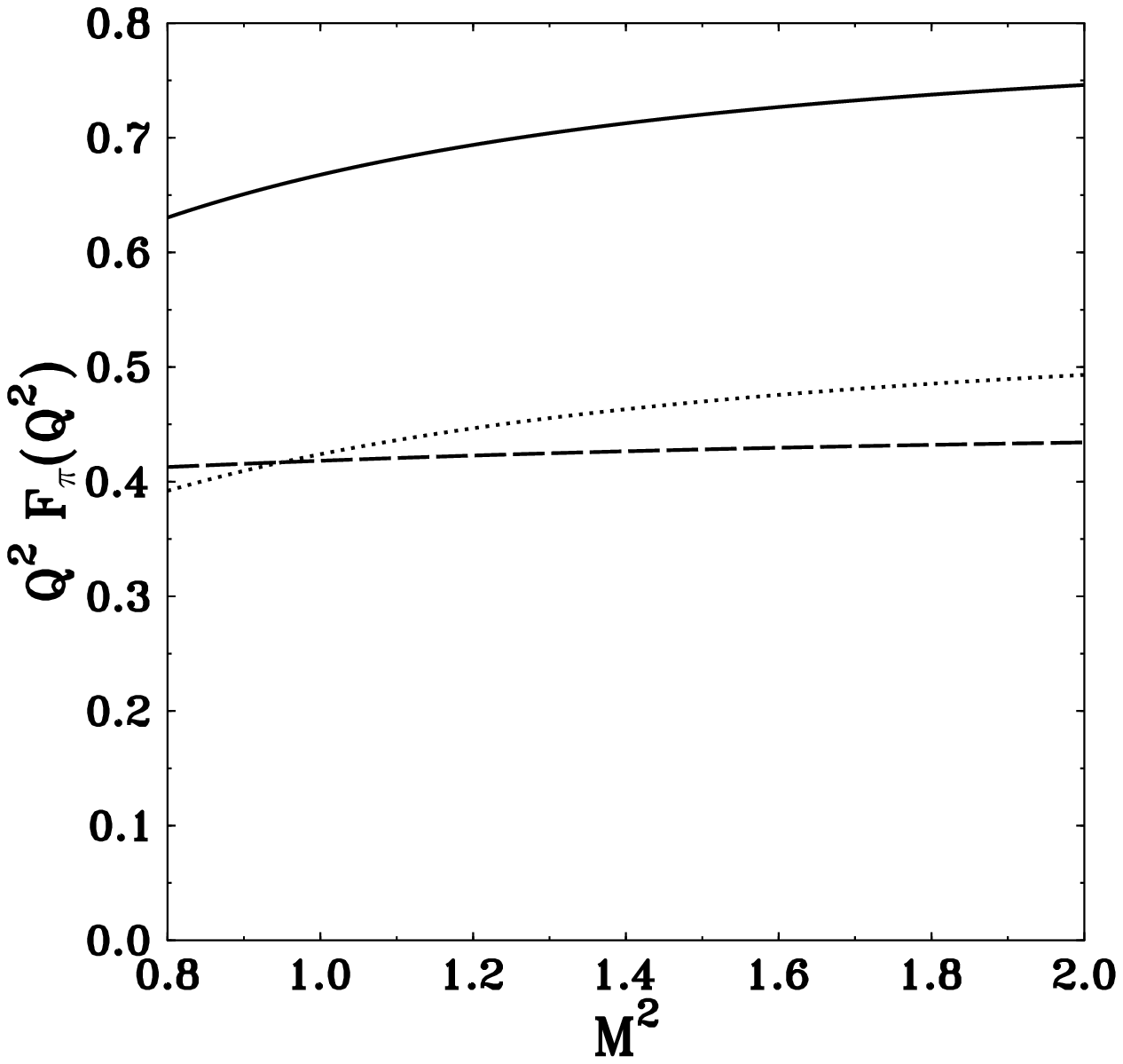,width=8cm}}
\caption
{\small The Borel parameter dependence of the light-cone sum rule for 
the asymptotic (left) and CZ (right) distribution amplitudes, 
at $Q^2$ =1, 3 and 10 GeV$^2$ shown by dashed, solid, and dotted
curves, respectively.    
}
\end{figure}
%%%%%%%%%%%%%%%%%% END FIGURE 8  %%%%%%%%%%
%  

There are several input parameters which should be specified
in the sum rule. First of all, the 
pion duality interval, $s_0=0.7$ GeV$^2$,
is determined by fitting the 2-point sum rule (\ref{3:SRfpi})  
to the pion decay constant $f_\pi=133 $ MeV. This sum rule 
is reliable for the corresponding Borel parameter 
$M^2_{2pt}= 0.7-1.0$ GeV$^2$. Having in mind that in the 
light-cone sum rule for the same pion channel the Borel 
parameter should be larger, typically 
of order $M^2 \simeq M^2_{2pt}/\langle u\rangle$,
we  assume $ 0.8 < M^2 < 1.5 $ GeV$^2$ as a fiducial interval.
We have checked that changing $s_0$ by $\pm 0.1$~GeV$^2$ does not
produce a significant effect, so that we stick to the above standard value 
\cite{SVZ,IofSmil,Rad82} in what follows.

The principal input is provided
by the leading twist distribution amplitude (see Appendix B) 
\begin{equation}
\varphi_\pi(u,\mu ) = 6 u \bar u \left[ 1 + 
a_2(\mu) C_2^{3/2}(u - \bar u) +  
a_4(\mu) C_4^{3/2}(u - \bar u) +\ldots \right]\,,
\label{phip}
\end{equation}
where the coefficients $a_n$ present the main 
nonperturbative input of interest.
%source of uncertainty
%and should be extracted, in principle, from experiment.
Taking into account poor accuracy of the present data as well
as considerable uncertainties in the sum rules themselves, we
cannot aim to distinguish between contributions of different Gegenbauer
polynomials. We put, therefore,  all $a_n, n=4,6,\ldots$ to zero and
consider the values $a_2=0$ (asymptotic distribution) and $a_2(1$~GeV$)=2/3$
(CZ distribution) as the two extreme alternatives. 
Calculations in this section are done taking into account 
the anomalous dimension  of $a_2$ to one-loop accuracy, see Appendix B.
Our main goal will be to determine $a_2$ from the comparison 
of the sum rule results with the experimental data.

The higher-twist distribution amplitudes and relevant parameters
represent another set of inputs. 
They are listed in the Appendix B. The
uncertainty in higher-twist corrections turns out to be sufficiently 
small and does not influence our final results.

Finally, one has to specify the
renormalization/factorization scale $\mu$. 
For this numerical analysis we use the $u$-dependent scale 
$\mu^2_u =(1-u)Q^2+u M^2$ for the leading twist 2 contributions
and simply take $\mu^2=M^2$ for the soft-region ($u>u_0$) 
dominated higher-twist corrections in the sum rule.  
The scale dependence is, in fact, rather mild.

The Borel parameter dependence of the 
sum rule is shown in Fig.~8 for three different values of $Q^2$. 
As can be seen from this figure,
the prediction for the form factor is sufficiently stable.
%
%
%
%%%%%%%%%%%%%%%%%% FIGURE 9  various contributions %%%%%%
\begin{figure}[t]
\centerline{\psfig{figure=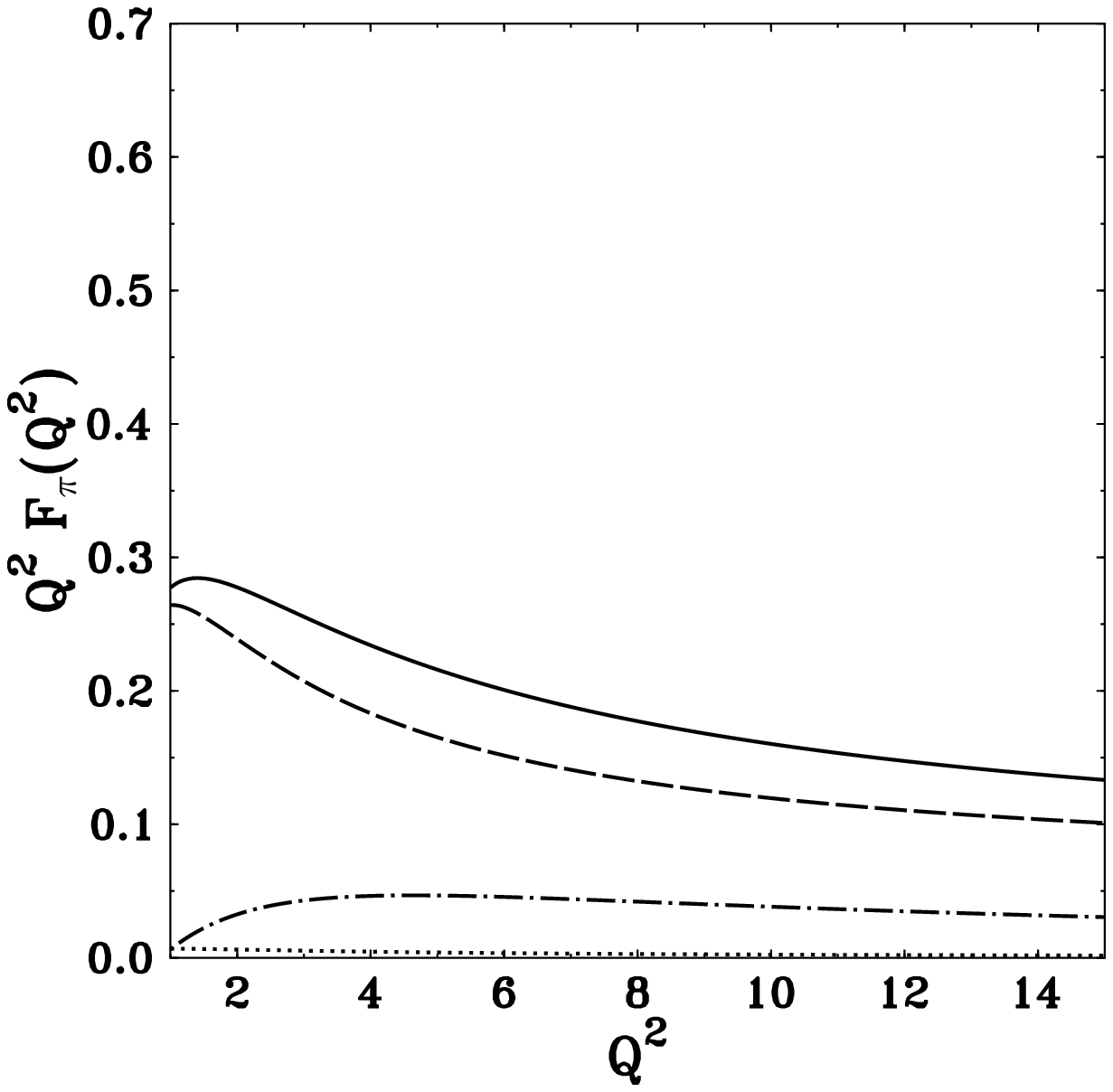,width=8cm}
            \psfig{figure=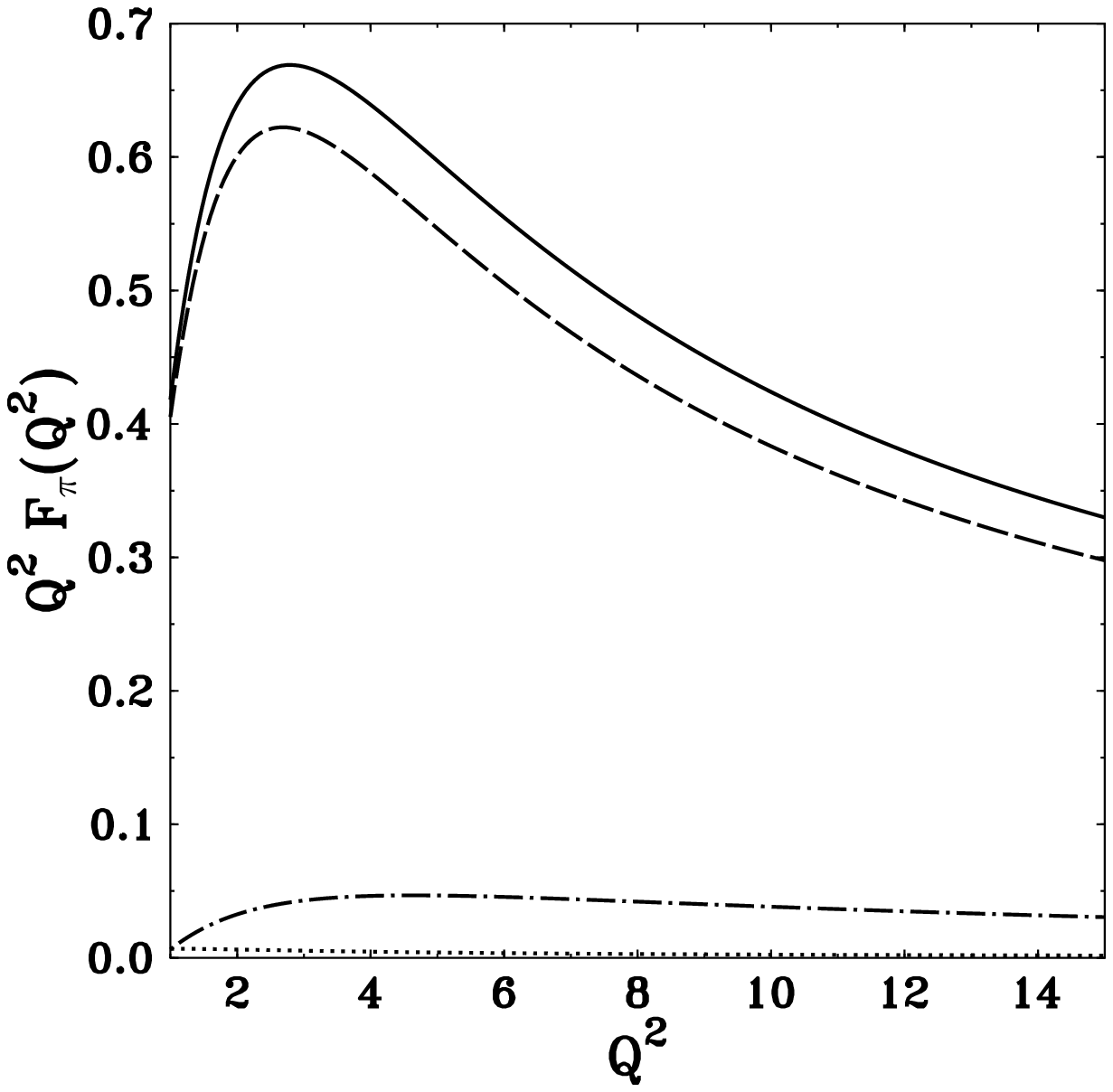,width=8cm}}
\caption
{\small The relative importance of contributions of different twist 
 in the light-cone sum rule
for the asymptotic (left) and CZ (right) 
twist 2 distribution amplitudes. Twist~2 (dashed), twist~4 (dot-dashed) 
and twist~6 (dotted) contributions and their sum (solid curve)
are plotted at $M^2 = 1$ GeV$^2$.}
\end{figure}
%%%%%%%%%%%%%%%%%%END FIGURE 9%%%%%%%%%%
%
%

The relative contributions of different twist to the sum rule
are shown as a function of $Q^2$ in Fig.~9. 
The twist~4 contribution 
does not exceed  25-30\% of the total result while the twist~6 
correction is negligibly small. This hierarchy  
reveals a good convergence of the light-cone expansion, at least 
at $Q^2 >1$~GeV$^2$. 
For lower values of $Q^2$ the higher-twist corrections 
become unstable and the approach breaks down.

Finally, in Fig. 10 we compare the light-cone sum rule 
calculation with the available experimental data in the interval 
$ 1 < Q^2 < 7 $~GeV$^2$ taken from \cite{Be78,Am86}. 
The dashed and dotted curves correspond to the asymptotic and 
CZ distribution amplitudes, respectively. The solid curve 
presents the best fit, yielding 
\begin{equation}
a_2^{\rm LCSR}(\mu=1\,\mbox{\rm GeV}) = 0.12\pm 0.07
\phantom{\,}^{+0.05}_{-0.07}
\label{fit:a2}
\end{equation}
with $ \chi^2 =13.9$ for 14 degrees of freedom. 
The first error comes from  the experimental uncertainty, whereas
the second error corresponds to the variation of the Borel parameter.
%
%
%%%%%%%%%%%%%%%%%%FIGURE 10 Pion Form Factor %%%%%%%%%%

\begin{figure}[t]
%\vspace{-1cm}
\centerline{\psfig{figure=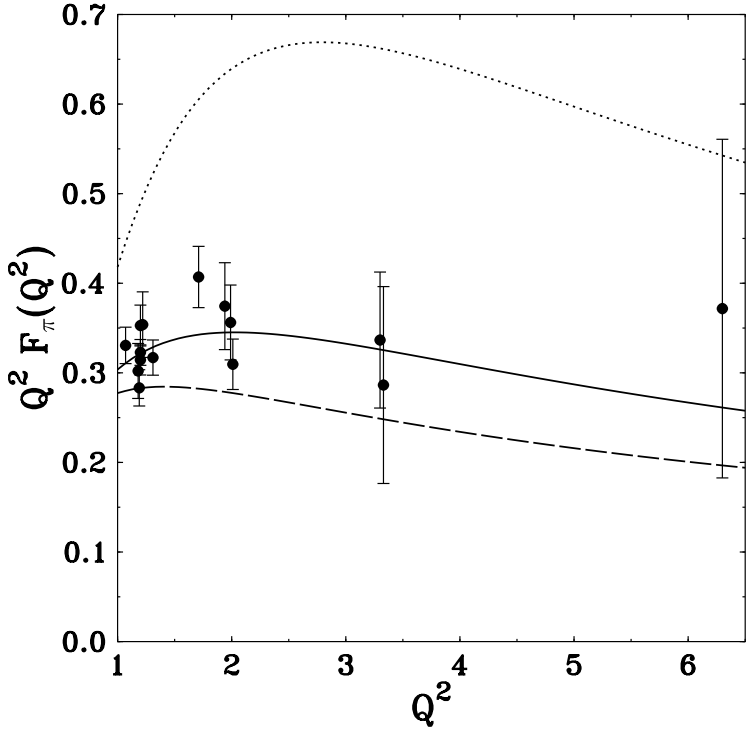,
%width=14cm}}
width=12cm}}
\caption
{\small The light-cone sum rule predictions for the pion electromagnetic 
  form factor using asymptotic distribution amplitude 
   (dashed), CZ distribution (dotted) and fit to the data (solid).  
}
\end{figure}
%%%%%%%%%%%%%END FIGURE 10 Pion Form Factor %%%%%%%%%%%
%
%
This value for the coefficient $a_2$ translates to the estimate
for the characteristic integral: 
\begin{equation}
 \int \! \frac{du}{\bar u}\, \varphi_\pi(u, \mu = 1\,\mbox{\rm GeV}) =
3.36\pm 0.21
\phantom{\,}^{+0.15}_{-0.21}\,.   
\end{equation}

\section{Matching with the NLO perturbative calculation}

As discussed in Sect.~3.3 the light-cone sum rule in the present form 
does not yet reproduce the full perturbative result in the asymptotic 
limit $Q^2\to\infty$. The missing terms correspond to higher-order 
corrections in the light-cone expansion and are difficult to calculate 
directly. Instead, one can make a matching of the light-cone sum rule 
to the NLO perturbative calculation by following the standard logic of 
the asymptotic expansion \cite{CSS}. 
To this end, write the twist 2 contribution to the sum rule defined 
in Eq.~(\ref{3:NLO}) as a sum of two terms, adding and subtracting 
the leading asymptotic expression at $Q^2\to \infty$ 
(see Eq.~(\ref{3:SRexpand}), first line)
\begin{eqnarray}
F^{(2)}_{\pi}(Q^2) &=& F^{(2)}_{\rm pert}(Q^2)+F^{(2)}_{\rm nonp}(Q^2)\,,
\nonumber\\
F^{(2)}_{\rm pert}(Q^2) &=&
\frac{\alpha_s}{2\pi}C_F\int\limits_0^{s_0} \frac{ds\,  e^{-s/{M^2}}}{Q^2} 
\int\limits_0^1 du\, \frac{\varphi_{\pi}(u)}{\bar u}\,, 
\nonumber\\
F^{(2)}_{\rm nonp}(Q^2) &=& F^{(2)}_{\pi}(Q^2)- F^{(2)}_{\rm pert}(Q^2)\,.
\end{eqnarray}
Following the arguments of \cite{exclusive} one can prove 
that higher-order corrections to the sum rule must assemble themselves 
to reproduce the factorized expression
\begin{eqnarray}
  F^{(2)}_{\rm pert}(Q^2) 
 &\Rightarrow& \int\limits_0^1 \!dx\int\limits_0^1\!dy\, 
  \varphi_\pi(x,\mu) T_H(x,y,Q^2,\mu)\, \varphi_\pi(y,\mu)
\label{6:improve}
\end{eqnarray}
with 
\begin{eqnarray}
 T_H(x,y,Q^2,\mu^2)&=& \frac{2\pi C_F\alpha_s(\mu)f_\pi^2}{N_cQ^2(1-x)(1-y)}
\left[1+\frac{\alpha_s(\mu)}{\pi}\,T_1(x,y,Q^2/\mu^2)+\ldots\right]\,. 
\end{eqnarray}
The remainder $F^{(2)}_{\rm nonp}(Q^2)$ is suppressed by a power 
of $Q^2$ and presents a true nonperturbative `higher twist' correction
to the usual perturbative result based on collinear factorization.

Making the substitution (\ref{6:improve}) we effectively take into account
all higher-order corrections to the sum rule to $O(1/Q^2)$ accuracy and 
neglect such corrections for power suppressed terms. 
This procedure tacitly implies that the 
numerical effect of the replacement (\ref{6:improve}) is more 
important than of uncalculated (higher-order and higher-twist) corrections 
to $F^{(2)}_{\rm nonp}(Q^2)$.
Such an assumption is natural, but in fact flawed because
of potential double counting of perturbative contributions of soft 
regions. As one signal for this problem, one may notice that 
the perturbative QCD expression 
suffers from infrared renormalons  in high orders \cite{renormalons}, which 
have to be cancelled by the corresponding renormalon contributions to 
$F^{(2)}_{\rm nonp}(Q^2)$. Using a full resummed expression for 
$F^{(2)}_{\rm pert}(Q^2)$ together with the leading-order expression for 
$F^{(2)}_{\rm nonp}(Q^2)$ destroys this intricate cancellation
and is, therefore, not fully consistent theoretically. 
This is a usual difficulty of making a separation between `perturbative'
and `nonperturbative' contributions, which has been discussed in much 
detail recently in context of the calculation of power corrections 
to deep inelastic scattering, Drell-Yan processes, event shapes in 
$e^+e^-$ annihilation and inclusive B-decays 
\cite{renormalons,hightwist}.

An alternative and theoretically better defined possibility is to 
make a separation between soft and hard contributions to the pion 
form factor with an 
explicit cutoff, as in Section 3, define $F^{(2)}_{\rm nonp}(Q^2)$
as the contribution coming from the soft region, and replace the 
`hard' contribution to the light-cone sum rule by the perturbative 
expression restricted to the same hard region. A difficulty in this case 
is that the soft-hard separation in the sum rule involves a cutoff in 
one momentum fraction only and becomes ambiguous when applied to
 the fully factorized expression (\ref{6:improve}) 
involving  two momentum fractions\footnote{A natural solution would be to 
introduce a cutoff in the transverse quark-antiquark separation rather 
than in the momentum fraction.}.

In the present paper we consider the first possibility because of 
its relative simplicity. We take into account
the radiative correction to the hard-scattering kernel 
\cite{FGOC81,DR81,BT87,MNP98} and the complete NLO evolution of the 
pion distribution amplitude \cite{Muller}, see Appendix B.  
To this accuracy
\begin{equation}
  F_{\rm pert}(Q^2) = F_{\rm pert}^{\rm LO} (Q^2) 
                     + F_{\rm pert}^{\rm NLO}(Q^2)\,, 
\end{equation}
where 
\begin{equation}
F_{\rm pert}^{\rm LO} (Q^2) = 
8 \pi\alpha_s(\mu^2) \frac{f_\pi^2}{Q^2}\Bigg | 1 + a_2^{LO}(\mu^2) 
                              + a_4^{LO}(\mu^2) \Bigg |^2
\end{equation}
and \cite{MNP98}
\begin{eqnarray}
F_{\rm pert}^{\rm NLO}(Q^2) &=&  \frac{16f_{\pi}^2}{Q^2}  
                \alpha_S^2(\nUV) 
                \Big[ 1 + a_2^{LO}(\nIR) + a_4^{LO}(\nIR) \Big]  
                \left[ a_2^{NLO}(\nIR) + a_4^{NLO}(\nIR) 
                         + \sum_{k=3}^{\infty} a_{2k}^{NLO}(\nIR) \right] 
\nonumber \\
&+&        8   \frac{f_{\pi}^2}{Q^2}  
                \alpha_S^2(\nUV) 
                \Bigg\{\frac{2}{3} 
                \left[ \frac{25}{6} a_2^{LO}(\nIR) + 
                       \frac{91}{15} a_4^{LO}(\nIR) \right]
                \Big[ 1 + a_2^{LO}(\nIR) + a_4^{LO}(\nIR) \Big]
                   \ln \frac{\nIR}{Q^2}
\nonumber \\
&+&
                   \frac{9}{4} 
                \Big[ 1 + a_2^{LO}(\nIR) + a_4^{LO}(\nIR) \Big] ^2
                   \!\!\ln \frac{\nUV}{Q^2} 
                   + 6.58 
                   + 24.99  a_2^{LO}(\nIR)+ 21.43  (a_2^{LO}(\nIR))^2 
 \nonumber \\ &+& 
                      32.81  a_4^{LO}(\nIR)+ 32.55  (a_4^{LO}(\nIR))^2
                   + 53.37  a_2^{LO}(\nIR)  a_4^{LO}(\nIR) \Bigg\}  .
\end{eqnarray}
Note that we do not distinguish between the renormalization and factorization
scales.

The complete expression for the form factor reads, respectively
\begin{equation}
  F_\pi(Q^2) = F_{\rm pert}(Q^2)+ 
                 F_{\rm nonp}^{(2)}(Q^2,M^2)+F_\pi^{(4)}(Q^2,M^2)
               + F_\pi^{(6)}(Q^2,M^2)\,,
\label{6:final}
\end{equation}
where we have taken into account that twist 4 and twist 6 corrections
to the light-cone sum rule receive no $1/Q^2$ contributions to our accuracy. 

For the numerical analysis, we still have to specify the factorization scale.
Since, after the subtraction of the asymptotic $1/Q^2$ contribution, 
the sum rule contribution is dominated by soft contributions, we 
choose  the fixed scale $\mu^2\sim M^2 = 1$ GeV$^2$ for simplicity.
For the perturbative contribution we use 
\begin{equation}
  \mu^2 = \kappa Q^2 + M^2,~~~ M^2 = 1\,\mbox{\rm GeV}^2 
\end{equation}
 with  parameter $\kappa$ in the range
\begin{equation}
1/4 < \kappa < 1\,. 
\end{equation}   
Note that with small values of $\kappa$ the scale is almost $Q^2$-independent.
Effectively, this choice amounts to doing the perturbative expansion to fixed 
(second) order and not attempting a renormalization group resummation.
This allows to minimize the problem with double counting of infrared
regions.

%%%%%%%%%%%%%%%%%% FIGURE 11  NLO  %%%%%%
\begin{figure}[t]
\centerline{\psfig{figure=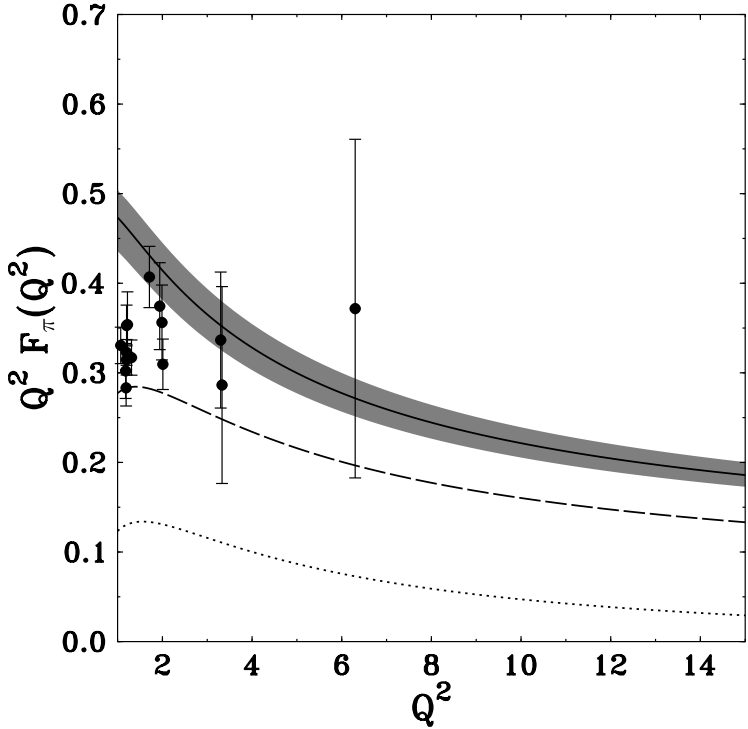,width=12cm}}
\caption
{\small 
%%%%%%%%%%%
The nonperturbative correction to the pion form factor (dotted) 
combined with the NLO perturbation theory (sum: solid curve),
compared with the pure light-cone sum rule result (dashed).
Asymptotic pion distribution amplitude is assumed.
The grey band shows the scale-dependence of the hard-scattering
contribution, see text. }
\end{figure}
%%%%%%%%%%%%%%%%%%END FIGURE 11%%%%%%%%%%
%

The numerical results are shown in Fig.~11 assuming the asymptotic pion 
distribution amplitude at the scale 1 GeV.
The result of the calculation using Eq.~(\ref{6:final}) and $\kappa=1/2$ 
is shown by the solid curve with the shaded band corresponding to variation 
of the scale parameter $\kappa$ in the given range. 
The dotted curve presents the nonperturbative 
contribution and the dashed curve is the  `pure' light-cone
sum rule calculation with the same parameters. The difference between the 
solid and the dashed curves presents, therefore, the net effect of the 
substitution (\ref{6:improve}).  

The nonperturbative (power suppressed) contribution to the pion form factor
shown by the dotted curve in Fig.~11 presents considerable interest
by itself. It is, obviously, independent on whether the substitution 
(\ref{6:improve}) is used (cf. discussion in the end of Sect.~3.3), 
and  turns out to be comfortably small. 
This smallness may appear to be unexpected after 
we have found large soft (end-point) corrections in Sect.~3, and is due 
to a  strong cancellation between the  leading order 
soft contribution to the sum rule (first line in Eq.~(\ref{3:expand}))
and the large radiative correction (second line in Eq.~(\ref{3:expand}))
corresponding to the sum of soft and hard contributions to $1/Q^4$
accuracy. As seen from  Eqs.~(\ref{3:localhard}), (\ref{3:localsoft}) 
the large negative hard contribution $\sim 1/Q^4$ 
plays the most important role in this cancellation.

Since, according to our analysis, the pion distribution amplitude
does not differ significantly from the asymptotic distribution,
the theoretical uncertainty in the light-cone sum rule calculation of 
the nonperturbative correction to the pion form factor is 
dominated by  dependence on the Borel parameter, as illustrated in Fig.~12.
%
%
%%%%%%%%%%%%%%%%%% FIGURE 12  nonperturbative  %%%%%%
\begin{figure}[h]
\centerline{\psfig{figure=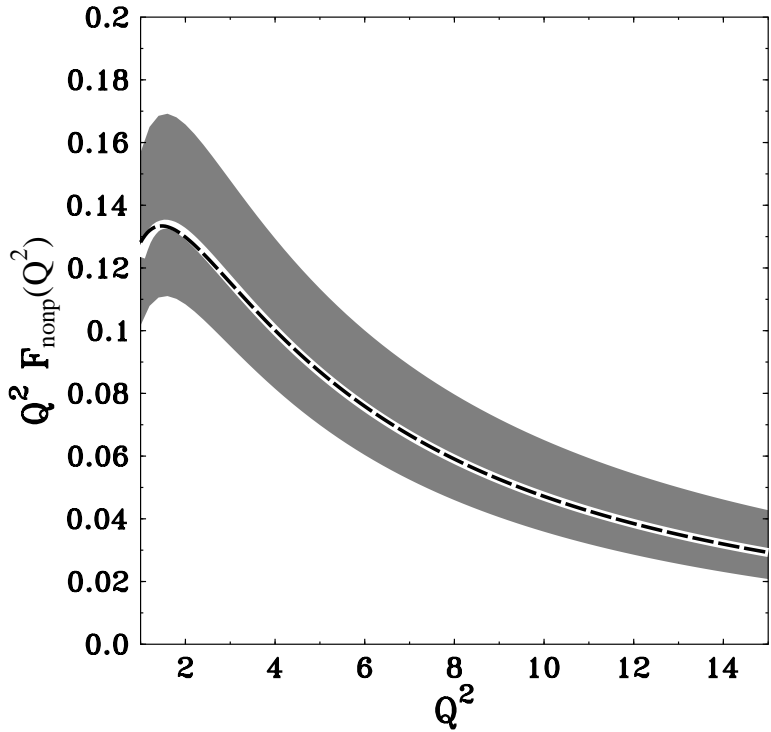,width=12cm}}
\caption
{\small 
%%%%%%%%%%%
The light-cone sum rule prediction for the nonperturbative correction
 to the pion form factor.
The grey band shows the sensitivity of the result to variation of the  
Borel parameter within $0.8 {\rm GeV}^2 < M^2 < 1.5 {\rm GeV}^2$.
The white line is the calculation for the standard
reference value $M^2 = 1 {\rm GeV}^2$ assumed throughout this paper
and the dashed curve is the fit (\protect\ref{fitnonp}). 
}
\end{figure}
%%%%%%%%%%%%%%%%%%END FIGURE 12%%%%%%%%%%
%
%
With the central values of parameters,  the nonperturbative 
correction can be parameterized in the region $1<Q^2<15$~ GeV$^2$ 
as\footnote{The given parametrisation should not be used for
larger values  of $Q^2$ since it has a wrong asymptotic behavior.}
\begin{equation}
 Q^2 F_{{\rm nonp}}(Q^2) = Q^2/(1.7046+1.0662 Q^2+0.0219 Q^4)^2
\label{fitnonp}
\end{equation}
(all numbers in GeV), and the theoretical error (the grey area in Fig.~12)
roughly corresponds to uncertainty in the overall normalization
of order $\pm 25\%$.

Choosing, as above, a model for the distribution amplitude at the scale 
1 GeV as a sum of the leading term and the second Gegenbauer polynomial,
and fitting the parameter $a_2(1$~GeV) to the data, we find:
\begin{eqnarray}
a_2^{}(\mu=1\,\mbox{\rm GeV}) = -0.06\pm 0.24
\pm 0.03\pm 0.03\,,
\nonumber\\
 \int \! \frac{du}{\bar u}\, \varphi_\pi(u, \mu = 1\,\mbox{\rm GeV}) =
2.82\pm 0.72\pm 0.09 \pm 0.09\,.
\label{fitnew:a2}
\end{eqnarray}
The first error comes from  the experimental uncertainty, 
the second error corresponds to uncertainty of the nonperturbative 
contribution (mainly dependence on the Borel parameter) and the 
third error is the scale dependence of the NLO perturbative result.
Combining the two estimates in Eqs.(\ref{fit:a2}) and (\ref{fitnew:a2}) 
and adding the errors in quadrature, we obtain as our final result
\begin{eqnarray}
a_2^{}(\mu=1\,\mbox{\rm GeV}) = 0.1\pm 0.1\,,
\nonumber\\
 \int \! \frac{du}{\bar u}\, \varphi_\pi(u, \mu = 1\,\mbox{\rm GeV}) =
3.3 \pm  0.3\,.
\label{final:a2}
\end{eqnarray}
This determination is dominated by the `pure' light-cone sum rule
result in which case we included the data points at lower values $Q^2$
having higher accuracy. The situation will change when sufficiently
precise data at $Q^2 > 2-3$~GeV$^2$ become available. In this region
the NLO perturbative prediction complemented by the `higher-twist'
power-suppressed correction in Eq.~(\ref{fitnonp}) becomes,
from our point of view,  a preferable
description, with potential theoretical accuracy of order 10\%. 
Note that the theoretical status of our result for the nonperturbative
(soft + hard) 
correction is similar to  model (or sum rule) determinations
of matrix elements of higher-twist operators in deep inelastic scattering.

\section{Conclusions}

Elaborating on the earlier proposal  \cite{BH94} we have given
in this paper a detailed quantitative analysis of the pion form factor
in the region of intermediate momentum transfers in the light-cone 
sum rule approach and also combining this technique with a complete
existing NLO perturbative calculation. Our results support the 
shape of the pion distribution amplitude that is close to 
the asymptotic expression and are inconsistent with  the CZ-type 
distributions. Our final estimate for the parameter $a_2$ characterizing
the deviation from the asymptotic form is given in Eq.~(\ref{final:a2})
\footnote{The smallness of nonasymptotic 
contributions to $\varphi_\pi$ is in agreement with the  
light-cone sum rule analysis \cite{AK} for the $\gamma^*\gamma \pi^0$ 
transition form factor, compared with the CLEO data \cite{CLEO}. For a 
recent update including NLO effects see \cite{SY}.}.   

Another important conclusion of our analysis is that the 
 nonperturbative contribution to the pion form factor turns out  to be 
rather moderate and does not exceed 30\% in the full $Q^2$ range,
see Fig.~11 and Fig.~12. One has to have in mind, however, that  separation
of `perturbative' and `nonperturbative' contributions is theoretically  not 
well defined because QCD perturbation theory is divergent \cite{renormalons}.
A fully theoretically consistent approach necessarily has to introduce 
an explicit scale separation, and in particular consider soft and
hard contributions to the pion form factor separately. 
We have  presented a detailed study of the 
soft-hard separation implemented with a hard momentum fraction cutoff
in Sect.~3. One finds that soft contributions are generally very large 
and the smallness of total nonperturbative correction is due to cancellations
between soft and hard terms of higher twist.
Thus, somewhat paradoxically, the nonperturbative effects in the 
pion form factor can be small and the soft contributions large, simultaneously!

To summarize, 
we believe that the light-cone sum rule approach presents a 
powerful and theoretically consistent framework to the analysis of hard 
exclusive reactions for intermediate momentum transfers.
Main and essential assumption of the method is duality, i.e that pion 
contribution can be isolated from the correlation function  by integrating the 
QCD spectral density in the certain energy range -- interval of duality.
While the numerical accuracy of this approximation can be disputed, 
it satisfies all known QCD constraints and provides a perfect laboratory
for the study of different interaction mechanisms involving several scales.
In particular, the scale-dependence of  the soft-hard 
separation studied in this work is of general validity.    

\vspace{0.8cm}

{\bf Acknowledgments}

The authors thank 
P.~Hoyer and  A.~ Radyushkin for useful discussions. 
M.M.~is grateful to
the Nordic Institute for Theoretical Physics (NORDITA) for hospitality.
A.K. acknowledges the NORDITA support during his visit
when this project was initiated. The work of A.K. is supported 
by the German Federal Ministry for Education and Research 
(BMBF) under contract number 05 7WZ91P (0).

\section*{Appendix A}
\app

Here we collect some useful formulae. 

The imaginary part of the radiative correction $H_1$
to the hard scattering amplitude reads:  
\begin{eqnarray}
\frac1{\pi}\mbox{Im} H_1(Q^2,s,u,\mu) 
=&& \Bigg(-9 + \frac{1}{3}\pi^2 + 3\ln[Q^2/\mu^2] - \ln[Q^2/\mu^2]^2 + 
3\ln\left[\frac{\bar u Q^2}{u\mu^2} \right]
\nonumber\\ 
 && -\ln^2\left[\frac{\bar u Q^2}{u \mu^2} \right]\Bigg )\delta(\rho)
\nonumber\\ 
&& + \Theta[-\rho]
\frac{Q^2(-2Q^2 + 3\rho + 5s + 2(Q^2 + \rho + s)\ln[-\rho/\mu^2]
)}{(Q^2 + s)^3\bar u u}
\nonumber \\
&&+2\Theta[-\rho]
\frac{Q^2(-(Q^2+s)\ln[s/\mu^2]
+ u(Q^2 - s + s\ln[s/\mu^2]))}{(Q^2 + s)^3\bar uu}
\nonumber \\
&&+2\Theta[\rho] \frac{Q^2(Q^2 - s + s\ln[s/\mu^2])}
{(Q^2 + s)^3\bar u}
\nonumber \\
&&
+\frac{Q^2s(-3 + 2\ln[s/\mu^2])}
    {(Q^2 + s)^2\bar u}\frac{d}{d\rho}(\ln[\rho/\mu^2]\Theta[\rho])
\nonumber \\
&&+2
\frac{Q^4\ln[s/\mu^2]}
    {(Q^2 + s)^2u}
\frac{d}{d\rho}(\ln[-\rho/\mu^2]\Theta[-\rho])
\nonumber \\
&&-2 \frac{Q^4}{(Q^2 + s)^2u}
\frac{d}{d\rho}(\ln^2[-\rho/\mu^2]\Theta[-\rho]) \,.
\end{eqnarray}
where $\rho = Q^2 \bar u - u s$.

The light-cone expansion of the quark propagator derived 
in (\cite{BB8889}): 
\begin{eqnarray}
S(x,0)&&\equiv -i\langle 0 | T \{q(x) \bar{q}(0)\}| 0 \rangle=
\nonumber
\\
&&
=\frac{\Gamma(d/2)\not\!x}{2\pi^2(-x^2)^{d/2}}  
+
\frac{\Gamma(d/2-1)}{16\pi^2(-x^2)^{d/2-1}}\int\limits_0^1~du 
\Big\{\bar{u}\not\!x \sigma_{\mu \nu} G^{\mu\nu}(ux) + 
u\sigma_{\mu \nu} G^{\mu\nu}(ux)\not\! x 
\nonumber
\\
&&
+
2i\bar{u}u\not\! x x_\rho D_\lambda G^{\rho\lambda}(ux) \Big\} 
- \frac{\Gamma(d/2-2)}{16\pi^2(-x^2)^{d/2-2}}
\int\limits_0^1 du\Big \{ i(\bar{u}u-\frac12)D_\mu G^{\mu\nu}(ux)\gamma_\nu  
\nonumber \\
&&
+\frac{i}{2} \bar{u}u(1-2u )x_\mu \not\!\! D D_\nu G^{\mu\nu}(ux) 
+\frac 12 \bar{u} u \epsilon_{\mu\nu\alpha\beta}x_\mu D^\alpha
D_\lambda G^{\lambda\beta}\gamma^{\nu}\gamma_5 \Big \}+ ...\,,
\end{eqnarray}
where $ G^{\mu\nu}= g_sG^{\mu\nu a}(\lambda^a/2)$, 
$Tr(\lambda^a\lambda^b) = 2\delta^{ab}$ 
and $d$ is the space-time dimension.
Only the terms proportional to the one gluon-field strength 
and its first covariant derivative are shown for brevity,

\appende

\section*{Appendix B}
\app
Here we define the light-cone distribution amplitudes
of the pion and specify their parameters.
The leading twist 2 amplitude $\varphi_\pi(u)$ and 
the twist~4 amplitudes $g_1(u)$ and $g_2(u)$  
enter  the light-cone expansion of the matrix element 
\begin{eqnarray}
\langle 0| \bar d(0) \gamma_{\mu} \gamma_5 u(x) | \pi^+(p) \rangle 
&=& i p_{\mu} f_\pi \int\limits_0^1 du e^{-iupx}
\left(\varphi_{\pi}(u) + x^2 g_1(u) \right)
\nonumber \\
&& + f_\pi\left(x_{\mu} - \frac{x^2 p_{\mu}}{px}\right) 
 \int\limits_0^1 du ~e^{-iupx} g_2(u) \;.
\end{eqnarray}
The QCD equations of motions relate $g_{1}$ and $g_2$ to 
the quark-antiquark-gluon twist~4 distributions
$\varphi_\parallel$, $\varphi_\perp$, $\tilde \varphi_\parallel$, 
and $\tilde \varphi_\perp$. The latter are defined by 
the following matrix elements
\cite{BF90}:
\begin{eqnarray}
\langle 0| \bar d(-x)\gamma_\mu \gamma_5 G_{\alpha\beta}(vx)
u(x) |\pi^+(p)\rangle &=& p_\mu \frac{p_\alpha x_\beta-p_\beta x_\alpha}{px}
f_\pi \int\! {\cal D}\alpha_i \varphi_\parallel(\alpha_i) e^{-ipx\tau(\alpha_i)}
\nonumber \\
&& + ( g^\perp_{\mu\alpha}p_\beta  - g^\perp_{\mu\beta}p_\alpha)
f_\pi \int\! {\cal D}\alpha_i \varphi_\perp(\alpha_i) e^{-ipx\tau(\alpha_i)}
,
\label{tw431}
\end{eqnarray}

\begin{eqnarray}
\langle 0| \bar d(-x)\gamma_\mu i\widetilde G_{\alpha\beta}(vx)
u(x) |\pi^+(p)\rangle &=& p_\mu \frac{p_\alpha x_\beta-p_\beta x_\alpha}{px}
f_\pi \int\! {\cal D}\alpha_i 
\tilde \varphi_\parallel(\alpha_i) e^{-ipx\tau(\alpha_i)}
\nonumber \\
&& + ( g^\perp_{\mu\alpha}p_\beta  - g^\perp_{\mu\beta}p_\alpha)
f_\pi \int\! {\cal D}\alpha_i 
\tilde\varphi_\perp(\alpha_i) e^{-ipx\tau(\alpha_i)},
\label{not}
\end{eqnarray}
where $\widetilde G_{\alpha\beta}= \frac 12 \epsilon_{\alpha\beta\rho\lambda}
G^{\rho\lambda}$ and the following  abbreviations are used: 
$$\tau(\alpha_i) = \alpha_1-\alpha_2+v\alpha_3,~~
{\cal D} \alpha_i = d\alpha_2d\alpha_2d\alpha_3
\delta \left(1- \alpha_1 - \alpha_2- \alpha_3 \right)$$
and 
$$
g_{\alpha\beta}^\perp = g_{\alpha\beta} - 
\frac{x_\alpha p_\beta + x_\beta p_\alpha}{px} ~\,.
$$

The distribution amplitudes are usually constructed 
using the formalism of the conformal expansion 
\cite{BF90}. To achieve a reasonable accuracy 
one tries to retain 
a few first terms of this expansion
in addition to the leading   
asymptotic term. The most familiar example is  
the  twist 2 pion distribution \cite{exclusive}
\begin{equation}
\varphi_\pi(u,\mu ) = 6 u \bar u \left[ 1 + 
a_2(\mu) C_2^{3/2}(u - \bar u) +  
a_4(\mu) C_4^{3/2}(u - \bar u)  +\ldots\right],
\label{phipi}
\end{equation}
where two orders of the conformal expansion 
in Gegenbauer polynomials $~C_{n}^{3/2}$
are explicitly shown, with
\begin{eqnarray}
C_2^{3/2}(x) &=& \frac{3}{2}(5 x^2 -1)\,,
\nonumber \\
C_4^{3/2}(x) &=& \frac{15}{8}(21x^4-14x^2 +1)\,.
\end{eqnarray}
The coefficients $a_n$ determine the nonasymptotic part of $\varphi_\pi$.
Their scale-dependence is given in the leading order  by
\begin{equation} 
a_n^{\rm LO}(\mu_2)=\left( 
\frac{\alpha_s(\mu_2)}{\alpha_s(\mu_1)}\right)^{-\gamma_n^{(0)}/\beta_0}
a_n^{\rm LO}(\mu_1)
\label{anom}
\end{equation}
where $\beta_0=11- \frac23 N_F$ and the anomalous dimensions are
\begin{equation}
\gamma_{n}^{(0)}=C_F\left[3 +\frac{2}{(n+1)(n+2)}-4\left(\sum_{k=1}^{n+1}
\frac1k\right)\right]\,.
\label{gamman}
\end{equation}
In the  numerical analysis in this paper we use, in particular,
 the asymptotic distribution 
(all $a_n=0$) and the CZ-distribution ($a_2(1.0~\mbox{GeV})= 2/3$, 
$a_{n>2}=0$). In next-to-leading order \cite{Muller}
the evolution requires an infinite sum
of coefficients:
\begin{eqnarray}
        a_2^{\rm NLO}(\nIR) & = &  
             a_2^{\rm LO}(\nIR)  P_{2}(\nIR)
             + Q_{20}(\nIR)
             \nonumber \\
        a_4^{\rm NLO}(\nIR) & = & 
             a_4^{\rm LO}(\nIR)  P_{4}(\nIR)
             + Q_{40}(\nIR) + a_2^{\rm LO}(\nIR)  Q_{42}(\nIR)              
             \nonumber \\
        a_{2k}^{\rm NLO}(\nIR) & = & 
             Q_{2k\;0}(\nIR) + a_2^{\rm LO}(\nIR)  Q_{2k\;2}(\nIR)
             + a_4^{\rm LO}(\nIR)  Q_{2k\;4}(\nIR) \,, ~k \geq 3 
\end{eqnarray}
with the following notations:
\begin{eqnarray}
  P_{k}(\nIR) & = & \frac{1}{4}  
              \left( \frac{\gamma_k^{(1)}}{2 \beta_0}
                 + \frac{\beta_1}{\beta_0^2} \gamma_k^{(0)} \right) 
              \left(1-\frac{\alpha_s(\nO)}{\alpha_s(\nIR)} \right), 
\nonumber \\
  Q_{kn}(\nIR) & = & \frac{(2k+3)}{(k+1)(k+2)} 
                \frac{(n+2)(n+1)}{2 (2n+3)}   
                C_{kn}^{(1)}  S_{kn}(\nIR) \,,
\nonumber \\
    S_{kn}(\nIR) & = &
 \frac{\gamma_k^{(0)}-\gamma_n^{(0)}}{\gamma_k^{(0)}-\gamma_n^{(0)}+\beta_0}
 \left[ 1 - \left( \frac{\alpha_s(\nO)}{\alpha_s(\nIR)}
          \right)^{1+(\gamma_k^{(0)}-\gamma_n^{(0)})/\beta_0} 
 \right] \,, 
\nonumber  \\
    C_{kn}^{(1)} & = & (2 n + 3)
 \left[ \frac{\gamma_n^{(0)}-\beta_0+4 C_F A_{kn}}{(k-n)(k+n+3)}
    + \frac{ 2 C_F (A_{kn} - \psi (k+2) + \psi(1) )}{(n+1)(n+2)} \right]\,,  
              \nonumber  \\ & & \nonumber  \\
    A_{kn} &=& \psi(\frac{k+n+4}{2})-\psi(\frac{k-n}{2})+
              2 \psi(k-n) - \psi(k+2)- \psi(1) \,, 
\end{eqnarray}
where
\begin{eqnarray}
    \psi(z) &=& \frac{d}{dz} (\ln \Gamma(z))\, ~~~~~ 
     \quad \beta_{1} = 102 - \frac{38}{3} N_F \,,
\nonumber \\
\gamma_0^{(1)} &=& 0,\quad \gamma_2^{(1)} = 111.03,\quad 
\gamma_4^{(1)} = 150.28\,.
\end{eqnarray}
Expressions for the twist~4 distributions including the next-to-leading 
corrections in conformal spin have been derived in \cite{BF90,Ball98}:
\begin{eqnarray}
g_1(u) &=& \frac{5}{2} \delta^2 \bar u^2 u + \frac{1}{2} \delta^2 \epsilon
\Bigg[ \bar u u(2 + 13\bar u u )
\nonumber\\
&&+ 10 u^3(2-3u+\frac{6}{5}u^2)\ln u + 10 
\bar u^3(2-3\bar u+\frac{6}{5}\bar u^2)\ln \bar u \Bigg],
\nonumber \\
g_2(u) &=& \frac{10}{3}\delta^2\bar u u(u-\bar u)\,,
\label{tw42p}
\end{eqnarray}
\begin{eqnarray}
\varphi_\parallel(\alpha_i) &=& 120\delta^2\epsilon
(\alpha_1-\alpha_2)
\alpha_1\alpha_2\alpha_3 \,,
\nonumber \\
\varphi_\perp(\alpha_i) &=& 30 \delta^2 (\alpha_1-\alpha_2)
\alpha_3^2 \left[
\frac{1}{3} +2\epsilon(1-2\alpha_3)\right] ,
\nonumber \\
\tilde \varphi_\parallel(\alpha_i) 
&=& - 120 \delta^2 \alpha_1\alpha_2 \alpha_3
\left[ \frac{1}{3} + \epsilon(1-3\alpha_3) \right],
\nonumber \\
\tilde \varphi_\perp(\alpha_i) &=& 30 \delta^2 \alpha_3^2
(1-\alpha_3)\left[\frac{1}{3} + 2 \epsilon(1-2\alpha_3)\right].
\label{tw43p}
\end{eqnarray}
To this accuracy, the specific combination 
(\ref{4:phi4}) of twist~4 distribution amplitudes reads:
\begin{eqnarray}
\varphi^{(4)}(u) &=& \frac{20}{3}\delta^2 u^2\bar{u}(3u-2)
 - 4 \delta^2 \epsilon u(2+11u-26u^2 +13u^3) 
\nonumber \\
&& -8 \delta^2 \epsilon \Bigg[ u^3(10-15u+6u^2)\ln(u)
  + \bar u^3 (1+3u+6u^2)\ln(1-u)\Bigg].
\end{eqnarray}
The normalizations of all these distributions 
are determined by a single nonperturbative parameter 
$\delta^2$ defined as 
\begin{equation}
\langle \pi |g_s\bar{d}\widetilde{G}_{\alpha\mu}\gamma^\alpha u|0 \rangle=
i\delta^2f_\pi q_\mu \,.
\label{delta1}
\end{equation}
The second parameter $\epsilon$ in Eqs.~(\ref{tw42p}) and (\ref{tw43p})  
is responsible for the first nonasymptotic corrections.
QCD sum rule estimates yield 
$\delta^2 \approx 0.2 {\rm GeV}^2$ \cite{Novikov:1984jt,Chernyak:1983is}:   
and $\epsilon \approx 0.5$ \cite{BF90} .
The scale-dependence of these parameters is given by \cite{BF90} 
\begin{eqnarray}
\delta^2(\mu_2^2) = \left(\frac{\alpha_s(\mu_2^2)}{\alpha_s(\mu_1^2)}
\right)^{32/(9\beta_0)}\delta^2(\mu_1^2)\,,
\nonumber
\\
(\delta^2\epsilon)(\mu_2^2) = \left(\frac{\alpha_s(\mu_2^2)}{\alpha_s(\mu_1^2)}
\right)^{10/\beta_0}(\delta^2\epsilon)(\mu_1^2)\,.
\end{eqnarray}
Finally, we should include in our list 
the twist~3 distribution amplitudes $\varphi_p$ and $\varphi_\sigma$ 
used in the calculation of the twist~6 corrections.
These distributions  parameterize the following matrix elements:
\begin{eqnarray}
\langle 0|\bar d(0) i\gamma_5 u(x)|\pi^+(p)\rangle &=& f_\pi \mu_\pi
\int\limits_0^1 \!du\, e^{-iupx} \varphi_p(u)\;,
\nonumber \\
\langle 0|\bar d(0) \sigma_{\alpha\beta}\gamma_5 u(x)|\pi^+(p)\rangle 
&=& \frac{i}{6}(p_\alpha x_\beta - p_\beta x_\alpha)
f_\pi \mu_\pi \int\limits_0^1 du\, e^{-iupx} \varphi_\sigma(u)\;,
\end{eqnarray}
where $\mu_\pi = m_\pi^2/(m_u+m_d)$.
The well known asymptotic form of these 
distributions:
\begin{equation}
\varphi_p (u) =1, \qquad \varphi_\sigma(u) = 6u\bar u \,, 
\end{equation}
is sufficient for the approximation adopted in this paper.
The relation (\ref{mupi}), together with the standard 
value of the quark condensate 
$\langle \bar{q}q \rangle(1\mbox{GeV})= (-240\mbox{MeV})^3$
yields  
$\mu_\pi(\mu = 1\mbox{GeV})\simeq 1.56$ GeV. Note that 
the normalization of the twist 6 correction is 
effectively determined
by the product $\alpha_s(\mu) \langle \bar{q}q \rangle^2(\mu)$
having in total a negligible anomalous dimension.

\appende

\appende

\end{document}